\documentclass[sigconf,screen,nonacm]{acmart}
\AtBeginDocument{%
  }

\usepackage{makecell}

\begin{document}

\title{Show me the evidence: Evaluating the role of evidence and natural language explanations in AI-supported fact-checking}

\author{Greta Warren}
\email{grwa@di.ku.dk}
\orcid{0000-0002-3804-2287}
\affiliation{%
  \institution{Department of Computer Science, University of Copenhagen}
  \city{Copenhagen}
  \country{Denmark}
}
\author{Jingyi Sun}
\email{jisu@di.ku.dk}
\orcid{0009-0003-2577-4627}
\affiliation{%
  \institution{Department of Computer Science, University of Copenhagen}
  \city{Copenhagen}
  \country{Denmark}
}
\author{Irina Shklovski}
\email{ias@di.ku.dk}
\orcid{0000-0003-1874-0958}
\affiliation{%
  \institution{University of Copenhagen}
  \city{Copenhagen}
  \country{Denmark}\\
    \institution{Linköping University}
  \city{Linköping}
  \country{Sweden}
}
\author{Isabelle Augenstein}
\email{augenstein@di.ku.dk}
\orcid{0000-0003-1562-7909}
\affiliation{%
  \institution{Department of Computer Science, University of Copenhagen}
  \city{Copenhagen}
  \country{Denmark}
}

\renewcommand{\shortauthors}{Warren et al.}

\begin{abstract}
    Although much research has focused on AI explanations to support decisions in complex 
    information-seeking tasks such as fact-checking, the role of evidence is surprisingly under-researched.
    In our study, we systematically varied explanation type, AI prediction certainty, and correctness of AI system advice for non-expert participants, who evaluated the veracity of claims and AI system predictions. Participants were provided the option of easily inspecting the underlying evidence. 
    We found that participants consistently relied on evidence to validate AI claims across all experimental conditions.
    When participants were presented with natural language explanations, evidence was used less frequently  although they relied on it when these explanations seemed insufficient or flawed.
    Qualitative data suggests that participants attempted to infer evidence source reliability, despite source identities being deliberately omitted.
    Our results demonstrate that evidence is a key ingredient in how people evaluate the reliability of information presented by an AI system and, in combination with natural language explanations, offers valuable support for decision-making.
    Further research is urgently needed to understand how evidence ought to be presented and how people engage with it in practice.

\end{abstract}

\ccsdesc[500]{Human-centered computing~Empirical studies in HCI}
\ccsdesc[300]{Human-centered computing~Empirical studies in collaborative and social computing}
\ccsdesc[300]{Computing methodologies~Natural language processing}

\keywords{explainable AI, fact-checking, explanation, natural language processing, misinformation}


\maketitle
\setcounter{page}{1}
\pagestyle{plain}

\section{Introduction}


Large Language Models (LLMs) are increasingly popular tools for decision support for high-stakes tasks that require retrieving and reasoning over complex information, such as fact-checking \cite{gallagher2024assessing,quelle2024perils,dierickx2024striking}.
However, these artificial intelligence (AI) models suffer from issues with factuality \cite{augenstein2024factualityLLMs, ji2023hallucination, dahl2024large} and bias \cite{gallegos-etal-2024-bias,marchiori-manerba-etal-2024-social} in their outputs, which means that relying on LLMs for decision-making is risky.
At the same time, the linguistic fluency and persuasive qualities of LLM-generated text \cite{pauli-etal-2025-measuring} can convince people to erroneously accept their outputs \cite{bender2021dangers}. 
As a result, overreliance, that is, accepting or following incorrect AI recommendations, is an increasingly recognised problem for human-AI interaction \cite{bansal2021teamperformance}, particularly in information-seeking and decision-support tasks \cite{kim2025fostering,budzyn2025endoscopist}.

Common mitigation approaches for overreliance include encouraging people to evaluate AI input before coming to a final decision \cite{bucinca2021cognitive,miller2023explainable} and providing explanations for AI decisions \cite{schemmer2023relianceexplanations}. 
A recent critique of explainable AI efforts highlights a lack of consideration for the importance of evidence in how people can integrate AI support into practice \cite{miller2023explainable}. 
As commercial search systems integrate LLM output, these services increasingly include both explanations and links to retrieved evidence to help people decide whether to rely on AI-generated search summaries or investigate further \cite{liu-etal-2023-evaluating,kim2025fostering}. 
Although source- or evidence-seeking behaviour is considered important to understanding and calibrating reliance on AI advice, surprisingly little research has addressed this issue directly \cite{ibrahim2025measuringmitigatingoverreliancenecessary}. 
Some LLM-assisted information-seeking studies have shown that including  explanations and links to evidence sources can help mitigate AI overreliance, although few people actively inspect the underlying evidence provided by the sources in the course of completing their tasks \cite{kim2024chatgptdissatisfaction,kim2025fostering}. Moreover, there is scant research on what may be more or less effective ways of presenting evidence for different types of information search and decision support activities. 
In particular, little prior work explores providing evidence as part of the experimental setup.
Making the evidence readily accessible alongside the AI system output decreases friction inherent in clicking through links to external information, and may enable people to more easily inspect the validity of the system's predictions, allowing for greater oversight and reliance calibration.


A second stream of overreliance mitigation approaches is the provision of explanations of AI system decisions, with the goal of supporting critical engagement with AI output, especially in decision-support tasks \cite{miller2019explanation,schemmer2023relianceexplanations,vasconcelos_explanations_2023}.
However, research on the relationship between explanations and reliance on AI system advice has produced conflicting findings. In some cases, explanations appear to enable more efficient use of AI support \cite{vasconcelos_explanations_2023}, although different types of explanations may also increase overreliance \cite{bansal2021teamperformance}. 
Communicating model uncertainty in AI outputs has been found to help people identify when advice may be unreliable \cite{liu2021ai,chiaburu2024uncertainty}, induce more deliberative thinking \cite{prabhudesai2023uncertainty}, and decrease overreliance \cite{kim2024LLMuncertainty}. While most have used numerical scores \cite{Kon_Aframework_MICCAI2024,papantonis2025not,warren2025explainablefactchecking}, or verbal hedges (e.g., ``I'm not sure but...'' \cite{kim2024LLMuncertainty}), to indicate model uncertainty, recent work suggests that people may find natural language explanations of uncertainty helpful and compelling \cite{sun2025clue,pmlr-v238-harsha-tanneru24a,steyvers2024calibrationgapmodelhuman}. Explanations of uncertainty may be an effective way to highlight inconsistencies in AI output, which can help mitigate overreliance \cite{si-etal-2024-verify-truthfulness,kim2025fostering}.

Building on this work, we sought to explore the roles of evidence and explanations in an AI-assisted information-seeking and decision-making task. We examined to what extent people used AI system explanations and the evidence underlying the AI system decision when evaluating an LLM-based AI system for fact-checking.
We conducted a mixed methods 3x2x2 controlled experiment (N=208) with crowdworkers to examine the effects of providing explanations and evidence in a fact-checking scenario and collected qualitative responses about how participants made their decisions. We compared two types of natural language explanations: standard verdict-focused explanations and explanations of AI uncertainty, along with a control condition consisting of only the typical numerical model certainty. We included the option to access the evidence documents used by the AI system in its output as part of the experimental setup across all conditions.
We measured which sources of information (e.g., evidence, explanations, own knowledge) participants found useful, their reliance on the AI system's advice, decision accuracy, subjective measures of confidence, usefulness and trust and qualitative evaluations of how they used the information available to them.

We found that participants overwhelmingly identified the evidence as the most useful information source, regardless of the correctness of AI advice, level of AI certainty, and whether they saw an explanation or not.
When participants were provided with natural language explanations, these were also viewed as helpful, particularly when the AI system's advice was correct or highly confident.
We also discovered an association between participants' self-reported trust in the AI system and the information they relied upon: those who placed more trust in the AI system tended to use AI explanations more, while those with lower levels of trust tended to rely more on the underlying evidence.
Together, our results highlight that making the underlying evidence for AI system outputs accessible has powerful effects for effective decision-support in information-seeking tasks.





\section{Related Work}

\subsection{Mitigating overreliance in decision-support systems}

Overreliance is typically defined as the acceptance of AI system output when it is incorrect, or delegation of decisions to an AI system when it is inappropriate to do so \cite{ibrahim2025measuringmitigatingoverreliancenecessary}, e.g., if a doctor were to wrongly accept an AI diagnosis without critically evaluating whether it aligned with the patient's symptoms.
Excessive deference to AI advice can lead to errors and severe harms, particularly in high-stakes scenarios such as clinical decision-support \cite{lee2023counterfactualclinical,kerasidou2022before}. The growing use of LLMs for decision-making and information-seeking in everyday life has also raised concerns about risks of professional deskilling \cite{bucinca2025decisionskills}, diminished educational outcomes \cite{zhai2024effects} and declines in general problem-solving and critical thinking abilities \cite{essel2024chatgpt}.
Approaches to mitigating overreliance on AI systems primarily fall into two categories: those that encourage users of AI systems to consider the \textit{evidence} for themselves (e.g., cognitive forcing \cite{bucinca2021cognitive} and evaluative AI paradigms \cite{miller2023explainable}), and those that encourage users to develop a better understanding of the AI system and its reliability, through providing \textit{explanations} (e.g., \cite{vasconcelos_explanations_2023}) and/or indications of model certainty (which may be numerical  \cite{prabhudesai2023uncertainty} or verbal \cite{kim2024LLMuncertainty}). In the following subsections, we examine prior work on both of these approaches.
Reliance is also related to trust, although they are distinct constructs  \cite{dzindolet2003role,hou2021aversionappreciation}; trust is typically construed as willingness or intent to rely on a third party or system \cite{kaplan2020trust}, while overreliance may be the behavioural outcome of placing too much trust in the system (although this may also stem from error or desire to eschew responsibility for a decision or outcome).
In this study, we measured the extent to which participants chose to use the AI system's predictions, to evaluate the impacts of presenting evidence and explanations for AI system outputs on participant's reliance. We also measured participants' subjective trust in the system to examine the relationship between trust and reliance on AI advice, explanations and evidence.

\subsection{The role of evidence in AI reliance}
A promising approach to reducing overreliance is including the evidence underpinning the AI system's prediction alongside its output.
Prior work has explored how decision-support paradigms can be used to promote cognitive engagement with the available evidence and encourage system users to come to their own conclusions.
Cognitive forcing is one such technique, in which the user of a decision-support system is presented with the AI system input and must make their own prediction as to what the outcome will be, before being shown the AI system's output \cite{bucinca2021cognitive}. 
The Evaluative AI paradigm is an alternative, which proposes that decision-support systems should provide evidence for and against each hypothetical outcome, rather than recommending a single prediction \cite{miller2023explainable}.
Recent work on reliance on LLMs for information-seeking and decision-support has focused on providing evidence in the form of clickable links to external information (e.g., news articles or government advice).
In a question-answering task, LLM responses that listed sources were associated with lower overreliance, and higher user confidence in their own answers \cite{kim2025fostering}.
However, the explanations provided by the AI system did not explain \textit{how or why} the sources cited support the AI system's prediction.
Moreover, participants only clicked on one or more of these sources 22-28\% of the time, suggesting that the positive impacts of sources may be more due to the credibility of the source linked (which were vetted by the authors to contain high-quality, correct information) than the information contained in the links themselves. While source credibility is a critical component of fact-checker decision-making \cite{liu2023checking}, it is notoriously difficult to integrate into automated decision-support systems because it tends to be highly variable and subjective \cite{andrews2025should}. 
In this study, we include evidence alongside all explanations, but omit source credibility information to limit impacts of familiarity and subjective bias. 
We hypothesised that evidence would be more helpful when explanations are not provided and could help mitigate overreliance.

\subsection{The role of explanations in AI reliance}
Explainable AI approaches aim to mitigate distrust in AI by helping people understand why particular decisions are made by the system \cite{shin2021effects,ferrario2022explainability}.
As such, the use of explainable AI to support high-stakes information-seeking tasks, such as fact-checking, is an area of considerable interest \cite{guo-etal-2022-survey}. 
Natural language explanations can also improve people's ability to detect untrustworthy news articles with equivalent accuracy to professional journalists \cite{schmitt2024explhuman}, while people were more accurate in assessing the credibility of news/social media posts when provided with text-based explanations compared to graphical explanations \cite{robbemond2022explmodality}.
However, LLM-generated explanations can also lead to overreliance on the AI system \cite{schmitt2024explhuman,si-etal-2024-verify-truthfulness}.
Prior work has examined how overreliance on AI system advice can be mitigated by communicating that the AI is uncertain in its prediction, for example, through visualisation \cite{prabhudesai2023uncertainty} or linguistic expressions of uncertainty such as prepending statements like ``I'm not sure, but...'' to LLM outputs \cite{kim2024LLMuncertainty}. 
Recent work suggests that natural language explanations can potentially enhance user understanding of model uncertainty \cite{steyvers2024calibrationgapmodelhuman}. For example, providing contrastive information to highlight inconsistency between explanations and model output can help users reason more critically about AI predictions \cite{si-etal-2024-verify-truthfulness}.
In this study we examine the utility of natural language explanations to support decision-making in automated fact-checking. Given that they can lead to overreliance, we evaluate two variants of natural language explanations -- a verdict-based explanation that follows the typical approach of justifying system predictions, and an uncertainty explanation designed to reduce overreliance by explaining the system's uncertainty in its prediction. 
We hypothesised that natural language explanations of uncertainty would be more useful than verdict based explanations in a fact-checking scenario and help mitigate overreliance by highlighting possible inconsistencies in AI predictions.


\subsection{AI decision-support for fact-checking}
We designed our experiment as a fact-checking task due to its relevance to both information-seeking and decision-support contexts, with potentially high stakes consequences, given that AI systems such as LLMs are increasingly used as sources of information and news \cite{simon2025generative}.
Recent work with fact-checking practitioners suggests that understanding how the AI system arrives at its prediction is key to effective use \cite{warren2025explainablefactchecking}.
The two most common approaches to explainability in this space are attribution-based explanations, which highlight the key tokens or words that contributed to the AI system's output \cite{popat2018declare}, and natural language explanations in the form of free-text justifications for the AI system's decision \cite{atanasova-etal-2020-generating-fact,kotonya-toni-2020-pubhealth}. 
While there is evidence that explanations can enhance understanding of AI systems by increasing model transparency \cite{eslami2019user}, research shows that they also can lead to problematic overreliance on AI advice \cite{bansal2021teamperformance,poursabzi-sangdeh-etal-2021-interpretability,zhang2020confidence,pafla2024LLMerrors}.
For instance, example-based and attribution-based explanations seem to have little effect on people's ability to correctly assess the veracity of claims \cite{linder2021level}, but can lead to reliance on AI system advice even where it can be demonstrably wrong \cite{lim2023xai}.
In this study, we designed a task in which participants are presented with the input (claim and evidence) and outputs of an AI system (predicted verdict, model uncertainty, AI explanation) and asked to decide whether to use the AI system's prediction or not, and to identify which information they used to make their own decision.

\section{Method}
\begin{figure*}[ht]
    \centering
    \includegraphics[width=\linewidth]{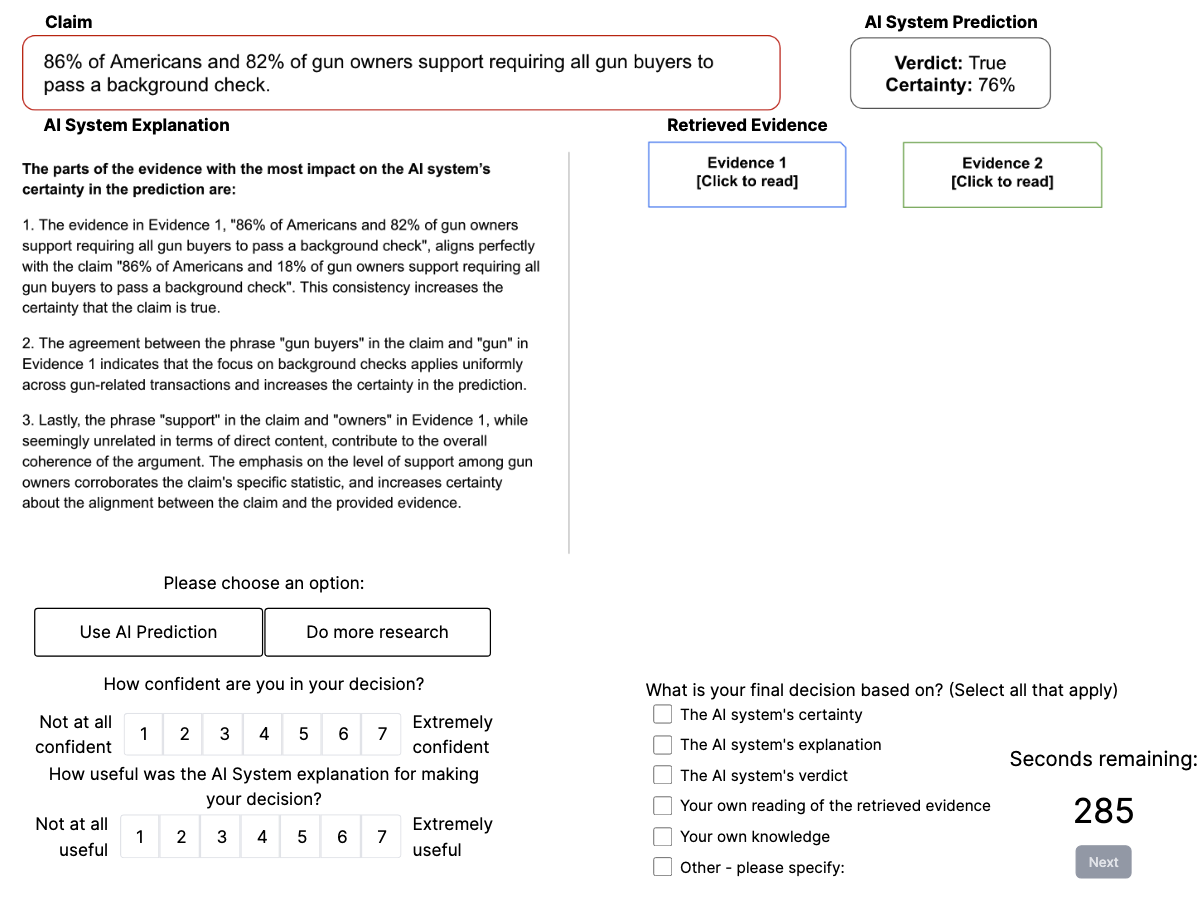}
    \caption{Example of task interface in Uncertainty Explanation condition}
    \label{fig:task_interface} 
\end{figure*}

We designed a controlled experiment to understand the role of different types of explanations and the impact of evidence evidence on how people make AI-supported decisions in a fact-checking scenario.
Participants were presented with the outputs of an AI system designed to support fact-checking (see Figure \ref{fig:task_interface} for an example of the interface) by predicting the veracity of claims. The system provided a numerical estimate of model certainty about the prediction, an explanation, and two evidence documents on which the prediction and explanation were based. 
The study was designed to assess the impact of explanation type on the fact-checking decision process, where participants could accept or decline the AI prediction. Each participant evaluated eight carefully selected combinations of claims and relevant evidence documents retrieved from online sources using one of three types of explanations (uncertainty explanations, verdict explanations, and no explanation) to support the process. We varied model certainty and model correctness counterbalanced within subjects. The study was approved by our institution's Research Ethics Committee.

\subsection{Experimental setup and design}

\subsubsection{Generating study materials}
The claims and evidence documents that formed the basis for the materials were drawn from the DRUID \cite{hagstrom2024realitycheckcontextutilisation} dataset, which consists of fact-checking claims drawn from professional fact-checking websites, such as Snopes (\url{https://www.snopes.com/}), and Full Fact (\url{https://fullfact.org/}), along with evidence documents retrieved from various online sources. 
For each claim and pair of evidence documents, we used Qwen2.5-14B-Instruct\footnote{\url{https://huggingface.co/Qwen/Qwen2.5-14B-Instruct}}, an open-weights, instruction-tuned language model with 14B parameters, to generate a predicted verdict and explanation for the given claim. We chose this model as a representative mid-sized open model that can be run locally (supporting reproducibility without relying on proprietary APIs), and that provides access to token-level logits required for our entropy-based uncertainty estimate. 
The AI system’s uncertainty was quantified by the entropy of its logits for generating the verdict, i.e., True or False (see Appendix \ref{app:uncertainty_est} for the estimation method).

%


\subsubsection{AI explanations}
We compared three types of explanations (see Table \ref{tab:example_expls} for examples). 
The \textit{Uncertainty explanations} provided reasons for the model certainty score. For example, conflicting evidence provided to the model may decrease the model's certainty that the predicted verdict is correct, while corroborating evidence may increase the model's certainty.
To generate the uncertainty explanations, we followed the CLUE method \cite{sun2025clue}. Given a claim and some evidence documents as input, this method first identifies key conflicting and concordant spans of text between the claim and evidence pieces that influence model certainty, and then instructs the model to describe how these conflicting or concordant span interactions, such as disagreements between two pieces of evidence, affect the certainty of the verdict prediction (see Appendix \ref{app:uncertainty_explanation_generation_prompt} for the instruction prompt).
The \textit{Verdict-focused explanations} were generated using a few-shot prompting method in which the model was instructed to explain why the verdict is predicted by referencing or summarising the provided evidence, and provided with three complete examples within the prompt (see Appendix \ref{app:generic_explanation_generation_prompt}). 
The `explanations' in the \textit{No explanation} condition were designed as a control condition and merely restated the AI system's verdict prediction and certainty score in text.

We assessed all generated explanations for quality, making minor edits to 
ensure consistency in structure (e.g., line-breaks and 2-3 numbered paragraphs for readability) and equivalent length in the explanations for the same claim. Where explanations did not explicitly explain the extracted text spans and the resulting AI system certainty percentage as instructed, we made minor edits to include this information to ensure that all explanations met our intended aims. We ensured that all Uncertainty explanations began with ``The parts of the evidence with the most impact on the AI system’s certainty in the prediction are:'' and that all Verdict explanations began with ``The parts of the evidence with the most impact on the AI system’s predicted verdict are:'' to ensure the target of each explanation type was clear.

\subsubsection{Interface design}
Following previous studies examining AI-supported fact-checking (e.g., \cite{schmitt2024explhuman,miranda2019automated, linder2021level}), we designed the interface to mimic a news dashboard.
Participants were presented with a fact-checking claim, the AI system's predicted verdict, certainty, and an explanation for the AI system's output. 
Participants could access the evidence by clicking on the evidence boxes in the interface to reveal them.
We chose to present the evidence within the experiment, rather than as links to external sources (as in e.g., \cite{kim2024llmsproducefaithfulexplanations,kim2025fostering}) for several reasons: 
(i) to mitigate the risk that evidence could be edited or removed, given that the AI outputs were pre-generated,
(ii) to maintain participants' focus on the experimental task, rather than becoming distracted by side-research, and hence (iii) to control the content, relevance, and length of information participants were provided, 
(iv) to allow us to measure the extent to which participants chose to view or hide the evidence,
and (v) to study the approach of providing evidence documents `in-window' used by a number of LLM-based chatbots designed specifically for knowledge intensive information-seeking tasks such as Consensus AI (\url{https://consensus.app/}) and Google Gemini's `Double-check response' feature (\url{https://gemini.google.com/app}).

\begin{table*}[h]
\renewcommand{\arraystretch}{1.2} 
\begin{tabular}{|p{3cm}|p{12cm}|}
\hline
\multicolumn{2}{|p{15cm}|}{\textbf{Claim:} A windmill could spin until it falls apart and never generate as much energy as was invested in building it.} \\ \hline
\multicolumn{2}{|p{15cm}|}{\textbf{Predicted Verdict:} False \quad \textbf{Certainty:} 31\%} \\ \hline
Uncertainty Explanation & 
The parts of the evidence with the most impact on the AI system’s certainty in the prediction are:

1. The evidence in Evidence 1, ``No, a professor didn’t say a windmill will `never generate' the energy invested in building it'', directly contradicts the claim’s assertion that windmills will never generate sufficient energy. This contradiction increases the AI system’s certainty that the claim is False.

2. The evidence in Evidence 1, ``will ‘never generate’ the energy invested'', aligns with the part of the claim ``spin until it falls apart and never generate as much energy as was invested'', reducing the AI system’s certainty that the claim is False.

3. The evidence in Evidence 1, ``expended to'', aligns with the phrase ``invested in building'' in the claim, indicating agreement on the energy expenditure aspect, reducing the certainty that the claim is False. \\ \hline

Verdict Explanation & 
The parts of the evidence with the most impact on the AI system’s predicted verdict are:

1. Evidence 1 explains that the claim comes from a misrepresented quote. The meme states that windmills ``will never generate as much energy as was expended to build the structures,'' but the original quote clarifies that ``while poorly placed windmills may never generate enough energy payback, a good wind site could generate it in three years or less.'' This shows the claim exaggerates and omits important context.

2. The same evidence highlights that well-sited windmills can generate enough energy to offset their construction cost quickly, making the claim that all windmills ``never generate'' enough energy inaccurate. The phrase ``a good wind site could generate it in three years or less'' directly contradicts the claim.

3. Evidence 2 reinforces this, stating that ``analyzing the energy used in the production of wind turbines does not show that it’s greater than the electricity produced over the turbine's working lifetime.'' This confirms that, over time, turbines produce more energy than was invested in building them.
 \\ \hline

No Explanation & The AI system has judged this claim to be False, with 31\% certainty. \\ \hline
\end{tabular}
\caption{Example of claim, corresponding AI prediction, certainty, and explanations}
\label{tab:example_expls}
\end{table*}

\subsection{Pre-testing experimental setup through think-aloud study}
We pre-tested our initial experimental setup with a think-aloud study to refine our design and materials and to observe in-situ how people reacted to the different types of explanations. We recruited five participants with diverse expertise
from our institution (T1-T3) and wider community (T4-T5) 
(see 
App. \ref{app:thinkaloud_ppts} for demographics). 
Sessions lasted approximately 1 hour, and were audio recorded and transcribed with participants' permission (see App. \ref{app:thinkaloud_protocol}). Each participant saw six claims, two with each type of explanation, in random order. We conducted semi-structured interviews upon task completion. 
We made several key design choices based on the observations and insights gathered during these sessions.


\subsubsection{Task Design.} Our initial study design asked participants to make a decision about whether the claim was true or false and gave no time limit for task completion.
After three think-aloud sessions, we observed that people often focused on reading the evidence, to the exclusion of the AI system decisions and explanations. A primary reason for this was that the decision about whether the claim was true or false did not necessarily require people to engage with the AI system, as the evidence was readily available and people could spend as much time as desired reading the evidence documents to form their decision.
Therefore, to encourage participants to engage with the AI system prediction, certainty and explanations, we made two main adaptations to the task.
Firstly, we altered the main decision participants made, from ``True'' vs ``False'' to ``Use AI Prediction'' vs ``Do more research''. In the experiment instructions (see Appendix \ref{app:instructs}), participants were told to imagine that their goal was to release assessments about whether a claim is true or false, and that their task was to decide whether to rely on the AI system prediction, or that more research was needed before releasing the assessment. The purpose of this change was to shift participants' focus towards evaluating the AI system and its usefulness, rather than merely evaluating the claim and the evidence.
Secondly, we introduced a time limit of 5 minutes per instance, to create a more realistic situation for decision-making, engender a greater sense of urgency and discourage participants from conducting side-research. When the time limit elapsed, evidence and explanations disappeared from the screen, and participants were prompted to make a final decision and move on to the next instance.
After introducing these changes, we conducted further think-aloud sessions with two additional participants, and observed that the changes that we made shifted participants' focus more towards the AI system. 

\subsubsection{Claims.} The think-aloud pre-testing also helped to identify potential issues with claims presented to participants. After each think-aloud session, we reviewed the comments participants made about the claims, and replaced claims that were too vague, too familiar, or too difficult to understand (e.g., ``I don't like the claim [...] it's very vague and I feel like it should just throw it out'' -- T1).

\subsubsection{Evidence sources.} Participants frequently commented on the importance of knowing the identity of sources: ``including the source of the evidence... It will allow me to know how trustworthy the evidence is, who made the claim'' (T2). While we wanted to avoid including specific names of sources to control for potential biases of participants \cite{andrews2025should,liao2014sourceexpertise}, we modified the task instructions to advise them that all evidence provided should be considered reasonably reliable.

\subsection{Controlled Experiment}

\subsubsection{Design and Measurements}
The final experimental design had a 3 (Explanation type: Uncertainty explanation vs Verdict explanation vs No explanation) x 2 (AI advice: correct vs incorrect) x 2 (AI certainty: high vs low) design to investigate the effects of explanations and evidence across different levels of AI correctness and AI certainty.
Explanation type was a between-participants factor, while AI correctness and AI uncertainty were within-participants factors.
In the main task (see Figure \ref{fig:task_interface}), participants were presented with a fact-checking claim, the AI system's predicted verdict, certainty, and an explanation for the AI system's output. Participants could access the evidence by clicking on the evidence boxes in the interface to reveal them. For each claim, participants were asked to decide whether to (a) use the AI system's prediction, or (b) do more research. If a participant chose to 'do more research', a follow-up question asked them to explain their decision given the options `The AI is incorrect', `More information is needed to make a decision', or `Other'.
For each decision, participants were asked to indicate what sources of information (i.e., AI verdict, AI certainty, AI explanation, evidence, their own knowledge, and other) their final decision was based on.
They were also asked to indicate (i) how confident they were in their own decision, and (ii) how useful the explanation was to their decision on a scale of 1-7. 
There was a time-limit of 5 minutes for each claim. If this elapsed, the evidence and AI explanations disappeared and participants were forced to make a decision and move on.
A post-task questionnaire collected subjective evaluations and free-text responses,
see Table \ref{tab:task_measurements} and Appendix \ref{app:subjective_scales} for details.

\begin{table*}[h!]
\renewcommand{\arraystretch}{1.3} 
\centering
\begin{tabular}{p{1.8cm} p{5cm} p{7.2cm}}
\toprule
\textbf{Task} & \textbf{Measurement} & \textbf{Response} \\
\midrule
\textbf{Main task} & AI advice use & Use AI prediction vs. Do more research \\
 & Information usage & Multiple choice: The AI system's verdict / AI certainty / AI explanation / Evidence / Own knowledge / Other \\
 & Decision confidence & 7-point scale: 1=``Not at all'' to 7=``Extremely'' \\
 & Explanation usefulness & 7-point scale: 1=``Not at all'' to 7=``Extremely'' \\
\midrule
\textbf{Post-task Questionnaire} & Information usage (qualitative) & Open text response to ``How did you use the information provided by the AI system to make your decisions?''\\
 & Explanation utility (qualitative)  & Open text response to ``Please explain why the AI system explanations were/weren't useful for determining how reliable the system was.'' \\
 & Explanation helpfulness (App. \ref{app:helpfulness}) 
 & 5-point scale: 1=``Not at all'' to 5=``Extremely'' \\
 & Trust Belief \& Intention \cite{mcknight2002impact} (App. \ref{app:trust_scales}) & 5-point scale: 1=``Strongly disagree'' to 5=``Strongly agree'' \\

 & Claim familiarity (App. \ref{app:claimfam}) & 4 options: No knowledge / Limited knowledge / Some knowledge / Full knowledge \\
\bottomrule
\end{tabular}
\caption{Overview of tasks, measurements, and response formats.}
\label{tab:task_measurements}
\end{table*}

\subsubsection{Participants}
A power analysis \cite{faul2009statistical} indicated that 207 participants were required for 90\% power for a medium effect (alpha<.05) for two-tailed tests.
Participants (N=208) were recruited using Prolific (\url{https://www.prolific.com/}), and randomly assigned to one of three conditions: Uncertainty explanation (n=70), Verdict-based explanation (n=69) and No explanation (n=69). Participant demographics are reported in Table \ref{tab:education}.
Participants were pre-screened to be native English speakers from Ireland, the United Kingdom, the United States, Australia, Canada and New Zealand, who had not participated in previous related studies. 
Twenty-two participants who failed more than one attention or memory check were excluded prior to data analysis. 
\begin{table}[h!]
    \centering
    \begin{tabular}{l|l c|l c}
    \hline
    \textbf{Age} & \textbf{Gender} & \textbf{\%} & \textbf{Education level} & \textbf{\%} \\ \hline
    M = 41.44  & Woman & 45.67 &   Doctorate degree & 5.28  \\
    SD = 13.34 & Man & 51.44  & Master's degree & 13.94  \\
     & Non-binary & 1.92 & Bachelor's degree & 35.58 \\
     & No answer & .96 & Some third-level & 24.52  \\
    & & &    High school graduate  & 17.79 \\
    & & &    Less than high school & 1.92 \\
    & & &    No answer  & 0.96 \\ \hline
    \end{tabular}
    \caption{Participant demographic details.}
    \label{tab:education}
\end{table}
\subsubsection{Materials}

Eight unique claims (and corresponding evidence documents) were selected from the DRUID \cite{hagstrom2024realitycheckcontextutilisation} dataset to balance (i) true and false verdicts, (ii) correct and incorrect AI predictions, (iii) high (>50\%) and low (<50\%) AI certainty. Explanations were consistent with the AI prediction, regardless of whether it was correct or incorrect.
For all claims, evidence and explanations see Appendix \ref{app:materials}.

\subsubsection{Procedure}
Participants received a link to the study hosted on Gorilla\footnote{\url{https://gorilla.sc/}} (see Appendix \ref{app:instructs} for task instructions). 
Participants read detailed instructions and completed a practice example.
They then progressed through the main task, providing responses for the eight claims in random order.
They then responded to a post-task questionnaire and optionally provided demographic information.
Participants were paid \pounds 4.50 for taking part. The average completion time was 29m 54s, corresponding to a payment rate of \pounds 9 per hour.

\subsubsection{Data analysis}
Quantitative analysis was performed using R \cite{R2021}.
Alongside traditional quantitative analysis, we analysed themes in our qualitative data from think-aloud interviews and free-text response questions. Two authors open-coded all of the data independently, periodically comparing codes and reconciling differences until achieving full agreement. Open codes were then combined into thematic categories to aid analysis \cite{braun2006using}. 

\section{Results}
The importance of evidence in our study was striking across quantitative and qualitative data. In contrast to prior research \cite{kim2025fostering}, 64\% of participants opened both evidence documents for every claim they assessed and just three (of 208) did not access either of the evidence documents. Participants reported finding evidence useful significantly more often than any other available information, regardless of condition. Evidence was also the primary topic of commentary in free-text responses, where participants explained which information they used to make decisions. Explanations were also judged to be helpful tools, particularly in identifying when the AI output was less reliable. In this section, we first discuss the original focus of the study, to assess the role of different types of explanations in claim assessment in a fact-checking scenario. We then discuss the role evidence played in this process and the implications of this finding. 

\subsection{The role of explanations in claim assessment}
Our initial hypothesis was that uncertainty explanations in natural language format would be more effective in supporting participants to evaluate whether to agree or disagree with the AI system. We analysed the extent to which participants used the AI system's prediction, depending on the explanation condition, correctness of the advice, and level of certainty of the system (see Table \ref{tab:evaluation_results}).
Predictably, participants were more likely to follow the AI system's advice when it was correct vs. incorrect, F(1, 207)=208.779, p <.001, $\eta_p^2$=.50, and when AI certainty was high vs. low, F(1, 207)=55.55, p <.001, $\eta_p^2$=.21.
Yet there was no difference in reliance based on the explanation condition F(2, 205)=1.825, p=.164, nor did this factor interact with advice correctness, F(2, 205)=0.1914, p=.9087, or certainty, F(2, 205)=0.5771, p=.749.
We did not observe any effect of explanation condition on confidence or trust judgments, nor did prior familiarity with claims affect results (see App. \ref{app:additional} for additional details).

\begin{table*}[h]
\centering
\begin{tabular}{lcccc}
\hline
\textbf{Measure} & 
\makecell{\textbf{All} \\ \textbf{conditions}} & 
\makecell{\textbf{Uncertainty} \\ \textbf{Explanation}} & 
\makecell{\textbf{Verdict} \\ \textbf{Explanation}} & 
\makecell{\textbf{No} \\ \textbf{Explanation}} \\
\hline
\multicolumn{5}{l}{\textbf{Use of AI prediction (\%)}} \\
Overall             & 54.56 & 51.79 & 57.79 & 54.17 \\
Correct AI advice   & 73.44 & 71.43 & 76.09 & 72.83 \\
Incorrect AI advice & 35.70 & 32.14 & 39.49 & 35.51 \\
High AI certainty   & 62.62 & 57.86 & 65.58 & 64.49 \\
Low AI certainty    & 46.51 & 45.71 & 50.00 & 43.84 \\

\multicolumn{5}{l}{\textbf{Explanation usefulness (1--7)}} \\
Overall             & 4.93 & 4.98 & 5.12 & 4.70 \\
Correct AI advice   & 5.30 & 5.28 & 5.47 & 5.16 \\
Incorrect AI advice & 4.56 & 4.68 & 4.76 & 4.24 \\
High AI certainty   & 5.22 & 5.15 & 5.43 & 5.07 \\
Low AI certainty    & 4.65 & 4.81 & 4.80 & 4.33 \\
\multicolumn{5}{l}{\textbf{Explanation helpfulness (1--5)}} \\
Overall             & 3.67 & 3.80 & 3.88 & 3.32 \\
\hline
\end{tabular}
\caption{Mean (i) use of AI prediction responses, (ii) explanation usefulness, and (iii) explanation helpfulness judgments by participants in each condition.}
\label{tab:evaluation_results}
\end{table*}

However, we observed a main effect of explanation condition on how often people used the explanations (see Table \ref{tab:information_use}), F(2, 205)=24.668, p<.001, $\eta_p^2$=.19. Post hoc Tukey HSD tests showed that participants in the two natural language explanation groups reported using the explanations in their decisions significantly more than the No Explanation group (Uncertainty: p<.001, d=.99, Verdict: p<.001, d=1.07). 
People judged that both Uncertainty (p=.003, d=.5) and Verdict (p=.013, d=.46) Explanations were more \textit{helpful} than No Explanation and that the Verdict Explanations were more \textit{useful} than No Explanation, p=.01, d=.34. Further, participants reported using the AI explanations more frequently when the AI system was correct (M=55.5\%) than when it was incorrect (M=48.1\%), F(1, 207)=12.125, p<.001, $\eta_p^2$=.055. No differences emerged between the two natural language explanation groups (p>.05 for all comparisons).

There was a main effect of explanation condition on how useful people judged the explanations, F(2, 205)=4.412, p=.013, $\eta_p^2$ =.02.
Bonferroni-corrected post hoc tests indicated that the Verdict Explanation group rated the explanations as more useful than No Explanation, p=.01, d=.34. 
Participants also judged the explanations to be more useful when the AI's advice was correct than when it was incorrect, F(1, 207)=81.53, p <.001, $\eta_p^2$ =.28, and when the AI system's certainty was high rather than low, F(1, 207)=66.868, p<.001, $\eta_p^2$=.24.

In the qualitative responses, participants complained about the lack of information in the No Explanation condition, although several mentioned using numerical uncertainty as a cue to be more sceptical. Participants in the natural language explanation conditions often found the explanations useful as summaries of evidence when they aligned with evidence. When explanations and evidence were misaligned, however, participants commented that this helped them understand the AI system's reasoning, aligning with prior research \cite{si-etal-2024-verify-truthfulness}: ``Sometimes by reading the explanations I could spot errors in the AI's logic.''  

Qualitative data also suggested reasons for why the natural language explanations did not differ along expected dimensions. Nearly all qualitative responses in the No Explanation condition mentioned using evidence to assess the claim, suggesting that having access to evidence documents compensated for  differences in explanation types. Participants in the Uncertainty condition had mixed opinions on the utility of these explanations. Some found them difficult to understand: ``on some of the choices the AI was very helpful but in others I found it very contradictory and confused.'' This suggests that more research on how to effectively explain uncertainty in natural language is necessary.  


\begin{table*}[h]
\centering
\begin{tabular}{lcccc}
\hline
\textbf{Measure} & 
\makecell{\textbf{All} \\ \textbf{conditions}} & 
\makecell{\textbf{Uncertainty} \\ \textbf{Explanation}} & 
\makecell{\textbf{Verdict} \\ \textbf{Explanation}} & 
\makecell{\textbf{No} \\ \textbf{Explanation}} \\
\hline
AI Verdict     & 41.10 & 40.00 & 45.83 & 37.50 \\
AI Explanation & 51.80 & 59.46 & 61.41 & 34.42 \\
AI Certainty   & 31.00 & 27.14 & 30.98 & 34.78 \\
Evidence       & 67.80 & 64.82 & 71.01 & 67.75 \\
Own knowledge  & 20.30 & 19.64 & 15.76 & 25.54 \\
Other          &  2.88 &  3.21 &  3.80 &  1.63 \\
\hline
\end{tabular}
\caption{Mean percentages of each source of information used.}
\label{tab:information_use}
\end{table*}

\subsection{The role of evidence in claim assessment}
Participants reported using the evidence most frequently (67.8\% overall), regardless of explanation condition, correctness of AI advice, or level of AI certainty.
A one-way repeated measures ANOVA indicated differences in how often participants used each information source in their decisions, F(5, 1035)=200.6, p<.001, $\eta_p^2$=.49.
Regarding the proportion of evidence documents opened, participants in the No Explanation group were more likely to open all evidence documents (83\%), compared to the explanation groups (55\% Uncertainty, p=.001, d =.62; 57\% Verdict, p=.003, d=.58).
This suggests that natural language explanations may often have offered sufficient information to make a judgment based on the explanation alone or a single evidence document.  
Mirroring the overall trend, participants in the Verdict-focused explanation condition relied more on the evidence than all other information sources, and relied on the explanations second most often, more so than the remaining information sources, p<.001 for all comparisons.
Participants in the Uncertainty explanation condition, however, reported using the evidence and explanations equally as often, p=.241. They relied on these two information sources significantly more than the others, p<.001 for all comparisons.
Participants in the No Explanation condition used evidence significantly more than any other information source, p <.001 for all comparisons. The `explanations' (restatements of AI verdict and certainty) were no more useful here than the AI verdict (p=.408), AI certainty (p =.911), or their own knowledge (p=.028; not significant on the Bonferroni-corrected alpha).

The qualitative data revealed interesting dynamics with respect to evidence. While evidence generally dominated as the most important information for evaluating claims, a common criticism was the absence of source information for the evidence documents. People commented that without source information they ``found it difficult to fully trust'' the evidence. In some cases, people tried to reverse engineer sources by paying attention to format, language, and, most often, the use of statistics. In fact, the use of statistics was at times perceived as more credible: ``if the evidence had actual numbers and data from respectable sources then usually I would side with the AI's verdict.'' Given that there was no source information in the evidence documents on whether the sources were ``respectable'' the credibility was clearly gained through the presence of numbers alone. This has implications for what kinds of evidence AI systems might present and suggests caution in how statistical evidence might be interpreted. 

\subsection{The role of subjective trust in reliance on AI advice, explanations, and evidence}

We investigated the relationships between participants' subjective trust judgments and their propensity to rely on (i) AI advice and use (ii) AI explanations and (iii) the underlying evidence in their decisions.
We found a moderate correlation between people's subjective trust in the AI system and their tendency to rely on the AI's recommendation (r=.425, p<.001), and trust predicted higher agreement with AI advice, $\beta$=0.91, SE=.135, p<.001, $R^2$=.181. 
Participants with higher trust were more likely to follow AI recommendations.
When we examined between group differences, we found that all explanation conditions exhibited a negative correlation between trust and AI agreement (Uncertainty: r=-0.35, p=0.003; Verdict: r=-0.17, p=0.176; No explanation: r=-0.35, p=0.003). 

Participants' trust in the AI system was also correlated with their tendency to use the AI explanations and underlying evidence in their decisions, although the effects were relatively small.
Our analysis showed a small positive correlation between participant's subjective trust judgments and the frequency with which they used the AI explanation to make their final decision (r=.26, p<.001), and trust predicted higher usage of AI explanations, $\beta$=0.838, SE=0.214, p<.001, $R^2$=.069.
On the other hand, trust in the AI system was negatively correlated with usage of evidence (r=-.29, p<.001), with trust predicting lower reliance on evidence, $\beta$=-1.015, SE=0.234, p<.001, $R^2$=.084. 
In other words, usage of evidence was associated with lower trust in the AI system, while usage of explanations was associated with higher trust.

Our qualitative findings corroborated the association between low trust in the AI system and reliance on evidence: ''I didn't trust it so had to check what it claimed''.
Participants also provided insight to how trust in the AI was shaped by how accurately the outcome reflected the evidence: ''I considered whether the AI's decision was based off the evidence, if their decision contradicted the evidence, this makes me feel that the AI was less trustworthy as it had not acknowledged the evidence properly''. 
People in the Explanation conditions also discussed using logical fallacies in explanations as cues to how trustworthy the system's decision was. Some participants in the Uncertainty condition reported using the sources of uncertainty to ''judge when to trust the system less''. 
This suggests that, when evidence is provided, explanations can play an important role in helping to determine whether the system has interpreted it correctly.

\section{Discussion}
Our results show that evidence is a key, yet under-researched component of how people make decisions when assessing AI system output, opening up important considerations for developing effective AI decision-support tools for information-seeking tasks. Strikingly, we observed that having access to evidence enabled participants to evaluate AI content successfully regardless of the type of explanation provided, although natural language explanations were appreciated. We also show that when people have access to evidence, natural language explanations do not necessarily to lead to overreliance (i.e., using an incorrect AI prediction) as previous studies have suggested \cite{he2025conversationalXAI,kim2025fostering}. Qualitative data suggest that access to evidence enables more critical assessment of AI output. 


\subsection{Availability is key for evidence engagement}
Previous work has shown that when AI systems provide people with links to external sources, relatively few (10-28\%) actually click the links \cite{kim2024LLMuncertainty,kim2025fostering}, and it remains unclear to what extent people meaningfully engage with the content of these sources. 
In contrast, approximately two-thirds of participants in the present study opened all of the available evidence documents and almost all opened at least one for each claim.
Several factors may account for this difference.
First, the evidence in our study was embedded directly within the task interface. Participants could click and read the evidence alongside the claim, AI prediction, and explanation, instead of following a link to an external webpage (e.g., \cite{kim2025fostering}), which likely reduced friction and effort associated with accessing the evidence.
Second, the explanations explicitly referenced the evidence, which may have encouraged participants to use it to verify the AI system's interpretation, or made it easier to identify the key parts of the evidence more efficiently. However, we observed that people reported using the evidence in their decisions regardless of whether explanations were provided.
It may also be the case that the fact-checking context of our task may have been perceived as more high-stakes than the question-answering setups in previous studies \cite{kim2024LLMuncertainty,kim2025fostering}, motivating participants to consider the information provided to them more carefully, though we note that click rates were extremely low ($\sim$7\%) even for potentially high-stakes health stimuli \cite{kim2024LLMuncertainty}.
These differences indicate that more research is needed to understand what kind of evidence is needed, in what context, and what are the best approaches to make it available to people using AI systems. 
Participants were also sensitive to inconsistencies between the verdict and the natural language explanation, but even where the explanation and the verdict aligned, inconsistencies between explanation and evidence alerted them to critically consider AI system output. As suggested by Kim et al. \cite{kim2025fostering}, the provision of sources themselves may also lead to more deliberative reasoning.

\subsection{Evidence-focused explanations may mitigate overreliance}
On one hand, our findings suggest a more optimistic state of affairs about people's reliance on and critical engagement with AI system explanations.
Prior studies in AI-supported fact-checking have highlighted that natural language explanations in the form of high-level abstractive summaries, designed to simulate how humans reason about claim veracity, can persuade and mislead people to incorrect conclusions \cite{schmitt2024explhuman,si-etal-2024-verify-truthfulness}.
However, the natural language explanations in our study, which extracted relevant parts of the evidence and explained how they contributed to the AI system's predicted verdict or certainty, did not lead participants to over-rely on incorrect advice in comparison to the No Explanation group.
This may indicate that explanations that focus on the evidence make it easier for people to identify inconsistencies, flawed argumentation or other cues that the explanation may be misleading. 
However, we also highlight potential for misuse of these explanations: cherry-picking evidence or otherwise using evidence out of context to falsely push a certain narrative could be highly persuasive. Although our qualitative data suggested that our participants were relatively alert to when this occurred, this may be less feasible when evidence documents are longer or more complex.
Future work in HCI is needed to examine how we can improve systems and explanations that allow people to detect when plausible-sounding explanations provide incorrect or misleading reasoning.
Further research should also examine evidence and explanation evaluation in more naturalistic, less structured task contexts and with domain experts who may interact differently with explanations than crowdworkers \cite{schmitt2024explhuman,szymanski2021visual}.


\subsection{Design implications for information-seeking \& decision-support systems}
The role of evidence in AI-assisted decision-making has received relatively little attention in the literature to date. There exists some theoretical research on the use of evidence in explanations, such as the evaluative AI framework \cite{miller2023explainable}, which builds on an earlier interpretability approach that utilised the information theory concept of weight of evidence \cite{melis2021human}. However, few studies have considered the role of evidence empirically or developed methods that use it along with AI predictions and advice. Many commercial LLM chatbots already provide people with sources and evidence documents alongside AI output; some automatically, while others require further inquiry about sources. However, there is currently little consensus on best practices. When current LLM implementations do give sources, LLMs struggle with factuality, i.e., reliably producing factual information \cite{augenstein2024factualityLLMs}, and often hallucinate references to external sources \cite{alkaissi2023artificial}, highlighting key gaps for further development of NLP and information retrieval tools to support AI-assisted fact-checking and other complex decision-making tasks.  Research in natural language explanations has largely focused on the production of fluent explanations (e.g., \cite{atanasova-etal-2020-generating-fact,kotonya2020fcpublichealth}) and how positively users evaluate them \cite{schlichtkrull2023usesfactchecking}.
Our findings call for efforts in NLP research to focus not only on how to produce explanations, but also consider the role of evidence in explaining model behaviour to the people who use them. Concurrently, HCI research is needed to gain fine-grained understanding of how evidence should be presented, and how evidence is evaluated and used by people in different AI-supported decision-making contexts.

\section{Conclusion}
We conducted a mixed methods experiment examining how evidence and explanations shape people's use of an LLM-based AI system for fact-checking.
We found that the evidence underpinning the AI output was the most important consideration for participants, regardless of whether explanatory information was provided.
Quantitative and qualitative data indicated that people found explanations to be useful indicators of where the AI system may be less reliable.
Our results highlight the central importance of evidence in AI-assisted information-seeking and decision support.
They also show how presenting evidence alongside AI outputs can substantially increase engagement with sources compared to current conventions of linking to external information.
Together, these findings suggest promising directions for developing best practices for mitigating overreliance through accessible evidence and evidence-focused explanations.

\begin{acks}
\noindent $\begin{array}{l}\includegraphics[width=1cm]{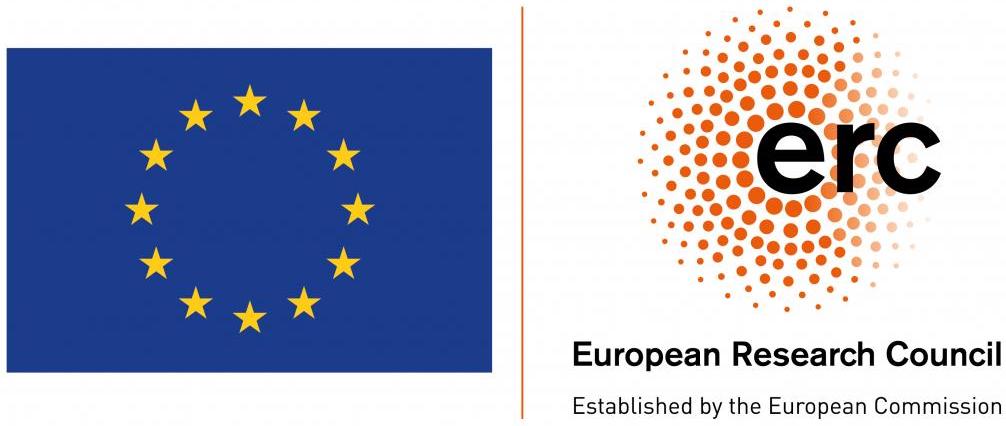} \end{array}$ This research was co-funded by the European Union (ERC, ExplainYourself, 101077481), and supported by the Pioneer Centre for AI, DNRF grant number P1. Views and opinions expressed are those of the authors only and do not necessarily reflect those of the European Union or the European Research Council. Neither the European Union nor the granting authority can be held responsible for them. 
\end{acks}

\bibliographystyle{ACM-Reference-Format}
\bibliography{2_references}

\appendix


\section{Think-Aloud Study Participants}
\label{app:thinkaloud_ppts}

\begin{table}[h]
    \centering
    \begin{tabular}{llll}
    \hline
      \textbf{ID}  &  \textbf{Age} & \textbf{Gender} & \textbf{Background} \\ \hline
      TI  &  24 & Female & Physics \\
      T2  &  58 & Female & International Relations \\
      T3  &  33 & Male & Computer Science \\
      T4  &  66 & Female & Translation \\
      T5  &  28 & Male & Sustainability \\
      \hline
    \end{tabular}
    \caption{Demographics of participants in think-aloud study}
    \label{tab:thinkaloudppts}
\end{table}

\section{Think-Aloud Study Protocol}
\label{app:thinkaloud_protocol}

\subsection*{Background knowledge}
\begin{itemize}
    \item How familiar are you with artificial intelligence or AI? Have you used any AI systems or LLMs like ChatGPT?
    \item How familiar are you with fact-checking?
\end{itemize}

\subsection*{Introduction to think-aloud study}

\begin{itemize}
    \item In this experiment, you’ll be testing an AI system designed to help with fact-checking. You’ll see some examples of claims that the AI system has been given, and its prediction about whether the claim is true or false. 
    \item You’ll see some more detailed instructions once you begin the study, and we’d like you to describe what you’re thinking as you read through the information you see, for example if something is unclear or confusing.
    \item Then, during the main study, as you’re evaluating the AI system and making decisions about the claims, we’d like you to also describe your impressions and thought processes, and talk through your decisions as you make them. At some points, I may ask you to explain a little more about why you did something.
    \item Then, at the end of the study I’ll ask you some questions about how you found the study overall.
    \end{itemize}

\subsection*{Think-aloud prompts (to be asked if participant doesn’t verbalise their thoughts)
}

\begin{itemize}
    \item What was your first impression when you saw this screen?
    \item Why did you check [the evidence/explanation/AI certainty]?
    \item How did you decide whether the claim was true or false? What did you base your decision on?
    \item Why were you confident (or not) in your decision?
    \item Why did you find the explanation (not) useful? What was (not) useful about it?
    \item Did the AI’s certainty score impact your decision in any way?
\end{itemize}

\subsection*{Post-study interview}

\begin{itemize}
    \item How did you find the task overall? Was anything unclear or confusing about what you were asked to do?
    \item Did you find anything difficult about the task?
    \item You saw different styles of explanation in the study. Were they helpful? Why/why not?
\end{itemize}

\section{Subjective Evaluation Scales} \label{app:subjective_scales}

\subsection{Explanation Helpfulness} \label{app:helpfulness}
To what extent were the AI system explanations helpful in determining how reliable the system was?

\noindent(1 = Not at all helpful,
2 = Somewhat helpful, 
3 = Neither, 
4 = Somewhat helpful, 
5 = Extremely helpful)

\subsection{Trust Belief and Intention Scales} \label{app:trust_scales}
On this page you are asked to provide your opinion on the AI system you used in the experiment.
Please select one answer for each statement
(1 = Strongly disagree,
2 = Somewhat disagree, 
3 = Neither, 
4 = Somewhat agree, 
5 = Strongly agree)

\noindent*Reverse-scored items

\begin{enumerate}
    \item The AI system is competent and effective in fact-checking claims
    \item Overall, the AI system is a capable and proficient information provider
    \item I would characterise the AI system as honest
    \item The AI system is NOT truthful in providing information to me*
    \item I believe that the AI system was developed to act in my (or the users') best interest
    \item The AI system was developed with good intentions to do its best to help me (or the users)
    \item When an important claim arises, I would NOT feel comfortable depending on the information provided by the AI system*
    \item I can always rely on the AI system to find information about claims
    \item I would feel comfortable acting on the information given to me by the AI system
    \item I would not hesitate to use the information the AI system supplied to me
\end{enumerate}

\subsection{Claim familiarity}\label{app:claimfam}

In the study, you saw 8 different real life claims, some of which you may have come across some of these claims before in your day-to-day life.

\noindent Please indicate whether you had read or come across any of these claims **before this experiment**, and if so, how much you knew about them.

\noindent\textbf{No knowledge} = I had never come across this claim before\\
\textbf{Limited knowledge} = I had not come across this specific claim before but I had previously come across something similar \\
\textbf{Some knowledge} = I had come across this claim before but I could not be sure that I knew whether it was true/false \\
\textbf{Full knowledge} = I had come across this claim and was confident that I knew it was true/false

\section{Additional analyses}\label{app:additional}
We collected data on how familiar participants were with the claims prior to the experiment, and on participants' confidence and trust judgements during the experiment. These variables did not appear to impact the main findings of the study, but we report them here for completeness and transparency.

\subsection{Claim familiarity}
On average, people were unfamiliar with the claims prior to the study 58\% of the time, had some knowledge 14\% of the time, had limited knowledge 24.94\% of the time, and had full knowledge of the claims 2.5\% of the time. 
People relied more on their own knowledge when they had some knowledge of the claim than when they did not, t(384)=6.87, p<.001, otherwise, no differences emerged in how people used the information available to them (p>.05 for all comparisons).
Verdict explanations were rated as more useful than no explanations when people had no familiarity with the claims, $\chi^2$(2)= 8.22, p=.016.
Otherwise, claim familiarity did not impact in how people judged the explanations, or how much they agreed with the AI system (p>.05 for all comparisons, see Table \ref{tab:familiarity_agreement_conf_use}).

\begin{table*}[h!]
\centering
\begin{tabular}{llcccc}
\hline
\textbf{Condition} & \textbf{Familiarity} & \textbf{n (\#responses)} & \textbf{Agreement (\%)} & \textbf{Confidence (1-7)} & \textbf{Usefulness (1-7)} \\
\hline
No Explanation     & Any knowledge & 230 & 53.5 & 5.78 & 4.83 \\
     & No knowledge & 322 & 55.7 & 5.79 & 4.60 \\
Verdict     & Any knowledge & 206 & 55.3 & 5.62 & 5.04 \\
     & No knowledge & 338 & 59.2 & 5.73 & 5.24 \\
Uncertainty & Any knowledge & 258 & 50.8 & 5.86 & 4.96 \\
 & No knowledge & 302  & 52.6 & 5.71 & 4.99 \\
\hline
\end{tabular}
\caption{Participant AI Agreement, confidence, and usefulness ratings for claims that people had either (a) any prior knowledge of or (b) no prior knowledge of, across explanation conditions.}
\label{tab:familiarity_agreement_conf_use}
\end{table*}



\subsection{Confidence and Subjective trust}

\subsubsection{Confidence}
There was no impact of explanation condition on how confident people were in their decisions, F(2, 205)=0.428, p=.652 (see Table \ref{tab:confidence_trust}).
Participants were more confident in their decisions when the AI's advice was correct than when it was incorrect, F(1, 207)=6.122, p=.014, $\eta_p^2$=.03, and when the AI system's certainty was high rather than low, F(1, 207)=27.886, p<.001, $\eta_p^2$=.12.

\begin{table}[h]
\centering
\begin{tabular}{lccc}
\hline
\textbf{Measure} & 
\makecell{\textbf{Uncertainty} \\ \textbf{Explanation}} & 
\makecell{\textbf{Verdict} \\ \textbf{Explanation}} & 
\makecell{\textbf{No} \\ \textbf{Explanation}} \\
\hline
\multicolumn{4}{l}{\textbf{Confidence (1--7)}} \\
Overall           & 5.78 & 5.70  & 5.78 \\
Correct AI advice    & 5.80 & 5.81 & 5.86 \\
Incorrect AI advice  & 5.76 & 5.59 & 5.70 \\
High AI certainty   & 5.87 & 5.88 & 5.88 \\
Low AI certainty    & 5.68 & 5.52 & 5.68  \\
\multicolumn{4}{l}{\textbf{Subjective trust (10---50)}} \\
Overall              & 33.4 & 33.9 & 32.0 \\
\hline
\end{tabular}
\caption{Mean confidence and trust judgments by participants in each condition.}
\label{tab:confidence_trust}
\end{table}

\subsubsection{Trust.}
The explanation groups did not differ in their judgments for the Trust scales, F(2, 205)=.091, p=0.40.

\section{Uncertainty Estimation and Explanation Generation Method}
\label{app:method_uncertainty_and_explanation}
\subsection{Uncertainty Estimation} \label{app:uncertainty_est}
By curating the instruction, we guided the model to generate the verdict along with the explanations, see Appendix \ref{app:generic_explanation_generation_prompt} and \ref{app:uncertainty_explanation_generation_prompt} for the prompts. Based on the model's verdict prediction, we derived its uncertainty as follows.
\label{app:uncertainty_score_generation_method}
To quantify model uncertainty for a verdict label on input $X$ (one claim plus two evidence passages), we compute the predictive entropy \citep{shannon1948,ulmer2022predictive} of the softmax distribution over logits, which requires only a single forward pass and is widely used. For candidate labels $\mathcal{Y}=\{\mathrm{True},\mathrm{False},\mathrm{NEI}\}$ with logits $\ell(y_i)$, the probability of label $y_i$ is
\begin{equation}
P(y_i \mid X) \;=\; \frac{\exp\!\big(\ell(y_i)\big)}{\sum_{j=1}^{|\mathcal{Y}|}\exp\!\big(\ell(y_j)\big)}.
\end{equation}
The uncertainty score for $X$ is the entropy of this distribution,
\begin{equation}
u(X) \;=\; -\sum_{y_i\in\mathcal{Y}} P(y_i \mid X)\,\log P(y_i \mid X),
\end{equation}

\subsection{Few-shot Prompt for Verdict Explanation Generation}
\label{app:generic_explanation_generation_prompt}
For the verdict explanation generation, we leveraged a three-shot prompt for instructing the model to generate the verdict along with the reasons towards this verdict, see Table \ref{tab:prompt_generic_explanation}.

\begin{table*}[t]
\setlength{\tabcolsep}{3pt}
\renewcommand{\arraystretch}{1.06}
\begin{tabular}{|p{2.8cm}|p{12.2cm}|}
\hline
\multicolumn{2}{|p{15cm}|}{\textbf{Task Description:} You are a helpful assistant. Here are your tasks: 
1. Supply the verdict for the claim based on the relationship between the claim and the two evidence passages. 
2. Explain your reasoning for the verdict prediction by identifying the three most influential span interactions from Claim-Evidence 1, Claim-Evidence 2, and Evidence 1-Evidence 2, and describing how each interaction (agree, disagree, or unrelated) affects your overall prediction. 
\textbf{Return format:} [ Verdict Prediction ] [ Explanation ]} \\ \hline

\textbf{SHOT 1 Input} &
\textbf{Claim:} Winston Churchill once stated, ``A pessimist sees the difficulty in every opportunity; an optimist sees the opportunity in every difficulty''. \\
& \textbf{Evidence 1:} \ldots ``A pessimist'', says the Rev. F. W. Cole, president of the Cardiff Free Church Council, \ldots \\
& \textbf{Evidence 2:} Winston Churchill's quote,``A pessimist sees the difficulty in every opportunity; an optimist sees the opportunity in every difficulty'', \ldots \\ \hline

\textbf{SHOT 1 Output} &
\textbf{Verdict Prediction:} False \quad \textbf{Explanation:} 
The parts of the evidence with the most impact on my predicted verdict are: The evidence in Evidence 1, ``says the Rev. F. W. Cole'', conflicts with the part of the claim ``Winston Churchill once stated''; the evidence in Evidence 2 that ``Winston Churchill's quote, `A pessimist sees the difficulty in every opportunity \ldots' '' aligns with the claim’s attribution; taken together, the attribution in Evidence 1 (``Rev. F. W. Cole'') versus Evidence 2 (``Winston Churchill'') is inconsistent, which weighs against the claim. \\ \hline

\textbf{SHOT 2 Input} &
\textbf{Claim:} During the last hundred years the temperature is increased about 0.1$^\circ$C because of carbon dioxide. The human contribution was about 0.01$^\circ$C. \\
& \textbf{Evidence 1:} ``The likely range of total human-caused global surface temperature increase \ldots\ is 0.8$^\circ$C to 1.3$^\circ$C (best estimate 1.07$^\circ$C [2.01 $^\circ$F]). Over this period, well-mixed greenhouse gases (GHGs) contributed a warming of 1.0$^\circ$C to 2.0$^\circ$C, and other human drivers (principally aerosols) contributed a cooling of 0.0$^\circ$C to 0.8$^\circ$C, natural (solar and volcanic) drivers changed global surface temperature by $-0.1^\circ$C to $+0.1^\circ$C, and internal variability changed it by $-0.2^\circ$C to $+0.2^\circ$C.'' \\
& \textbf{Evidence 2:} ``After using a bunch of nonsense calculations to say that 0.1 $^\circ$C warming is from CO2, they say that 90\% of the change in CO2 is caused by the oceans. \ldots\ the oceans are actually absorbing CO2 \ldots\ their conclusions cannot be supported.'' \\ \hline

\textbf{SHOT 2 Output} &
\textbf{Verdict Prediction:} False \quad \textbf{Explanation:} 
The parts of the evidence with the most impact on my predicted verdict are: Evidence 1 reports human-caused warming near 1$^\circ$C overall and 1.0--2.0$^\circ$C from well-mixed greenhouse gases, contradicting the claim’s ``about 0.1$^\circ$C because of carbon dioxide'' and ``about 0.01$^\circ$C'' human contribution; Evidence 2 challenges the assertion that ``0.1 $^\circ$C warming is from CO2'', which also goes against the claim; the segment in Evidence 1 about aerosol cooling and natural drivers ($-0.1^\circ$C to $+0.1^\circ$C) is unrelated to Evidence 2’s point about oceans absorbing CO2. \\ \hline

\textbf{SHOT 3 Input} &
\textbf{Claim:} Vitamin D appears increase COVID-19 mortality rates \\
& \textbf{Evidence 1:} If vitamin D does in fact reduce the severity of COVID-19 in regard to pneumonia/ARDS, inflammation, inflammatory cytokines and thrombosis, \ldots\ supplements would offer a relatively easy option to decrease the impact of the pandemic. \\
& \textbf{Evidence 2:} A Vitamin D deficiency that is not sufficiently treated is associated with COVID-19 risk. \\ \hline

\textbf{SHOT 3 Output} &
\textbf{Verdict Prediction:} False \quad \textbf{Explanation:} 
The parts of the evidence with the most impact on my predicted verdict are: Evidence 1 indicates that vitamin D reduces COVID-19 severity, which opposes the claim that vitamin D increases mortality; Evidence 2 links untreated vitamin D deficiency with higher COVID-19 risk, which also runs against the claim; Evidence 1's indication of benefit from adequate vitamin D is consistent with Evidence 2’s association between deficiency and risk, jointly weighing against the claim. \\ \hline

\textbf{NEW INSTANCE} &
\textbf{Claim:} \{ CLAIM \} \textbf{Evidence 1:} \{ E1 \} \textbf{Evidence 2:} \{ E2 \}  \textbf{Your answer:} \\ \hline


\end{tabular}
\caption{Verdict explanation generation prompt}
\label{tab:prompt_generic_explanation}
\end{table*}

\subsection{Few-shot Prompt for Uncertainty Explanation Generation}
\label{app:uncertainty_explanation_generation_prompt}
For the uncertainty explanation generation, we used a three-shot prompt to instruct the model to generate the verdict, along with the reasons to forward its verdict prediction certainty by referring to the key span interactions extracted in the previous step, following \cite{sun2025clue}, see the prompts in Table \ref{tab:prompt_uncertainty_explanation}.

\begin{table*}[t]
\renewcommand{\arraystretch}{1.0}
\begin{tabular}{|p{3cm}|p{12cm}|}
\hline
\multicolumn{2}{|p{15cm}|}{\textbf{Task Description:} You are a helpful assistant. Here are your tasks : 
1. Supply the verdict for the claim based on the relationship between the claim and the two evidence passages. 
2. Explain your certainty for your verdict prediction by referring to the three span interactions provided below ( Claim-Evidence 1 , Claim-Evidence 2 , Evidence 1-Evidence 2) and describing how each interaction’s relation (agree, disagree, or unrelated) affects your overall certainty . 
\textbf{Return format:} [ Verdict Prediction ] [ Explanation ]} \\ \hline

\textbf{SHOT 1 Input} &
\textbf{Claim:} Winston Churchill once stated, ``A pessimist sees the difficulty in every opportunity; an optimist sees the opportunity in every difficulty''. \\
& \textbf{Evidence 1:} \ldots\ ``\,A pessimist'', says the Rev. F. W. Cole, president of the Cardiff Free Church Council, \ldots \\
& \textbf{Evidence 2:} Winston Churchill's quote, ``A pessimist sees the difficulty in every opportunity; an optimist sees the opportunity in every difficulty'', \ldots \\
& \textbf{Span interactions (Pre-filled):} 
(1) ``Winston Churchill once stated'' - ``says the Rev. F. W. Cole'' (C - E1) relation: [disagree]; 
(2) ``Winston Churchill once stated'' - ``Winston Churchill's quote, `A pessimist sees the difficulty in every opportunity' '' (C - E2) relation: [agree]; 
(3) `` `A pessimist', says the Rev. F. W. Cole, president of the Cardiff Free Church Council'' - ``Winston Churchill's quote, `A pessimist sees the difficulty in every opportunity' '' (E1 - E2) relation: [disagree] \\ \hline

\textbf{SHOT 1 Output} &
\textbf{Verdict Prediction:} False \quad \textbf{Uncertainty Explanation:}
The parts of the evidence with the most impact on my certainty in the prediction are: the claim's phrase ``Winston Churchill once stated'' conflicts with Evidence 1’s attribution to ``the Rev. F. W. Cole'', which decreases my certainty in the claim; the same claim wording matches Evidence 2, which introduces the line as ``Winston Churchill’s quote'', increasing my certainty; since one passage attributes the quote to Rev. Cole and the other to Churchill, the evidences disagree with each other, and that inconsistency lowers my overall certainty and supports predicting False. \\ \hline

\textbf{SHOT 2 Input} &
\textbf{Claim:} During the last hundred years the temperature is increased about 0.1$^\circ$C because of carbon dioxide. The human contribution was about 0.01$^\circ$C. \\
& \textbf{Evidence 1:} \ldots\ well-mixed greenhouse gases (GHGs) contributed a warming of 1.0$^\circ$C to 2.0$^\circ$C \ldots \\
& \textbf{Evidence 2:} \ldots\ ``0.1 $^\circ$C warming is from CO2 \ldots\ cannot be supported''. \ldots \\
& \textbf{Span interactions (Pre-filled):} 
(1) ``temperature is increased about 0.1$^\circ$C because of carbon dioxide'' - ``well-mixed greenhouse gases (GHGs) contributed a warming of 1.0$^\circ$C to 2.0$^\circ$C'' (C - E1) relation: [disagree]; 
(2) ``0.1$^\circ$C warming is from CO2'' - ``0.1 $^\circ$C warming is from CO2 \ldots\ cannot be supported'' (C - E2) relation: [disagree]; 
(3) ``well-mixed greenhouse gases \ldots\ 1.0$^\circ$C to 2.0$^\circ$C'' - ``0.1 $^\circ$C warming is from CO2 \ldots\ cannot be supported'' (E1 - E2) relation: [agree] \\ \hline

\textbf{SHOT 2 Output} &
\textbf{Verdict Prediction:} False \quad \textbf{Uncertainty Explanation:}
The parts of the evidence with the most impact on my certainty in the prediction are: the claim's ``about 0.1$^\circ$C because of carbon dioxide'' contrasts with Evidence 1’s ``1.0$^\circ$C to 2.0$^\circ$C'' contribution from well‑mixed greenhouse gases, which pushes my certainty toward False; Evidence 2 explicitly states that ``0.1 $^\circ$C warming is from CO2 ... cannot be supported'', further moving my certainty toward False; together, the larger range in Evidence 1 and the rejection of the 0.1 $^\circ$C figure in Evidence 2 reinforce each other, increasing my certainty in the False prediction. \\ \hline

\multicolumn{2}{|p{15cm}|}{\emph{SHOT 3 omitted for space (structure identical to SHOT 1 and SHOT 2).}} \\ \hline

\textbf{NEW INSTANCE} &
\textbf{Claim:} \{ CLAIM \} \\
& \textbf{Evidence 1:} \{ E1 \} \\
& \textbf{Evidence 2:} \{ E2 \} \\
& \textbf{Span interactions (Pre-filled):} 
(1) `\{ SPAN1 - A \}' - `\{ SPAN1 - B \}' (C - E1) relation: \{ REL1 \}; 
(2) `\{ SPAN2 - A \}' - `\{ SPAN2 - B \}' (C - E2) relation: \{ REL2 \}; 
(3) `\{ SPAN3 - A \}' - `\{ SPAN3 - B \}' (E1 - E2) relation: \{ REL3 \} \\ 
& \textbf{Your answer:} \\ \hline

\end{tabular}
\caption{Uncertainty explanation generation prompt}
\label{tab:prompt_uncertainty_explanation}
\end{table*}

\section{Task Instructions}
\label{app:instructs}

\subsection*{Screen 1}
\textbf{Welcome to the experiment!}\\
\textbf{What is this study about?}\\
This study is about AI systems and fact-checking.
Fact-checking is the process of verifying the truthfulness of claims, statements, or stories by comparing them against credible evidence and sources.

\noindent In this study, you will see examples of an AI system designed to help people with fact-checking. Your task will be to carefully read and evaluate the information provided by the AI system.

\noindent You will see and evaluate 8 different claims in total. 
After you have evaluated these claims, you will answer some questions about your experiences of and opinions on the AI system and demographics.

\noindent You will have limited time to evaluate each claim. The main study will take 25-30 minutes to complete, while the questionnaire will take 5-10 minutes.

\noindent Before starting, make sure your browser window is set to full screen. Please make sure you are sitting somewhere comfortable and quiet so that you can complete the experiment in one sitting!

\noindent Click `Next' to read the instructions for the task.

\subsection*{Screen 2}
\textbf{What do I have to do?} \\
Imagine your job is to release assessments about whether a claim is true or false. You will see claims, an AI system’s prediction about whether this claim is true or false, how certain the system is about its label, and an explanation of the AI system’s prediction. These explanations are intended to help you decide how to interpret the AI prediction. You have the option to see the evidence that the AI system used to make its prediction. Your task is to decide whether you can rely on the AI system prediction or need to do more research before releasing the assessment of the claim. \\

\noindent\textbf{Important: }
Please only consider the provided information — claim, verdict, uncertainty, explanations \& evidence documents (if applicable) — when evaluating explanations. Sometimes you will be familiar with the claim, but we ask you to approach each claim as new, whether or not you have seen it before. It doesn’t matter whether you personally agree or disagree with the claim or evidence. We are asking you to evaluate what the AI produces: if you were to see this claim for the first time, would you find the explanation provided by the AI useful? Would you be willing to rely on the AI verdict to make a statement about the claim? 

\subsection*{Screen 3}
\textbf{About the AI System}\\
The AI System you will evaluate in this study is based on a Large Language Model (LLM).
This AI System has been designed to help people with fact-checking.
It has been trained to predict whether the claim is true or false based on the evidence it has been provided.
The AI System also provides explanations for its outputs, which you will evaluate in this study.

\noindent \textbf{What information will I see? (Definitions)}

\begin{itemize}
    \item A \textbf{claim} is some statement about the world. It may be true, false, or somewhere in between.
    \item \textbf{Evidence} is additional information is typically necessary to verify the truthfulness of a claim. An evidence document consists of one or several sentences extracted from an external source for the particular claim. In this study, you will have the option of reading two evidence document excerpts that have been retrieved for a claim. These evidence document excerpts may or may not agree with each other. Please assume that all evidence documents are reasonably reliable. 
    \item A \textbf{verdict} is reached based on the available evidence, regarding whether a claim is true or false.
    \item A \textbf{certainty score} indicates how confident an AI system about the correctness of its prediction. For example, a Certainty score of 99\% indicates very high confidence that the predicted verdict is correct, but a score of 1\% indicates very low confidence that the predicted verdict is correct. 
    \item An \textbf{explanation} for the AI system’s prediction explains the sources of uncertainty regarding the verdict. In other words, the explanation highlights parts of the evidence that increase or reduce its certainty that its verdict is correct.
    \item Your \textbf{TASK} is to decide whether you can rely on the verdict about the claim produced by the AI system, based on the information available, and how confident you are in your judgment, and decide how helpful the explanation is in making your decision. 
\end{itemize}

\noindent On the next page, you will see an example of the task.

\subsection*{Screen 4}

\begin{figure}[ht]
    \centering
    \includegraphics[width=.9\linewidth]{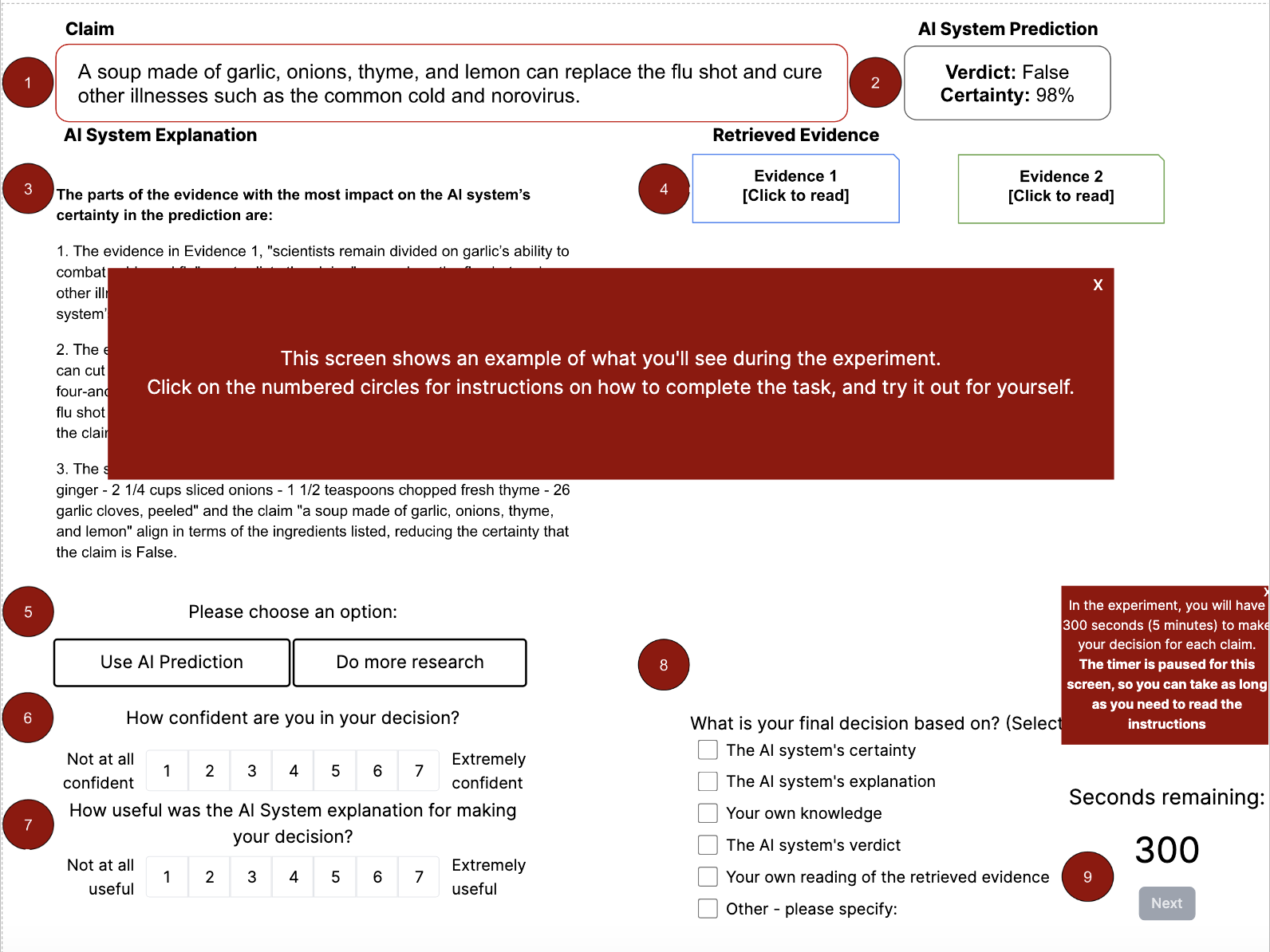}
    \label{fig:app:instruct_example}
\end{figure}

\begin{enumerate}
    \item First, read the claim that the AI system has been given.
    \item Next, read the verdict output by the AI system, along with its certainty that the output is correct.
    \item Next, read the explanation for the AI system's output. Here,  the explanation highlights parts of the evidence that increase or reduce its certainty that its verdict is correct.
    \item If you want to, you can check the evidence that the AI system used to make its prediction, by clicking on `Evidence 1' and/or `Evidence 2'.
    \item Based on the information provided, decide whether you can rely on the verdict produced by the AI system. Try to disregard any prior knowledge or opinions you may have and focus only on the information presented here. Based on the information provided, decide whether you can rely on the verdict produced by the AI system. Try to disregard any prior knowledge or opinions you may have and focus only on the information presented here.
    \item Indicate how confident you are about your decision whether to rely on the AI system verdict or not.
    \item Rate how useful the AI system explanation above was for making your decision. 
    \item Select the sources of information that you based you final decision on. There is no limit to how many you can select. If you used other sources not listed here, you can type them in `Other'.
    \item When you have completed all these steps, press `Next' to move on to the next claim.
You have 300 seconds (5 minutes) to complete your decision for each claim. If the timer runs out before you finish, your response will not be counted.
\end{enumerate}

\subsection*{Screen 5}
Thank you for completing the practice example!\\
\noindent Now, you're ready to start the study. You will see 8 different claims in total.\\
\noindent When you're ready, click `Next' to begin the study.

\section{Study Materials}
\label{app:materials}

\subsection{Claim 1 (True, Correct, High Certainty)}
\textbf{Claim:} 86\% of Americans and 82\% of gun owners support requiring all gun buyers to pass a background check.

\noindent \textbf{AI Prediction:} True

\noindent \textbf{AI Certainty:} 76\%

\subsubsection*{Evidence 1}
Statistic: 86\% of Americans and 82\% of gun owners support requiring all gun buyers to pass a background check, no matter where they buy the gun and no matter who they buy it from''. [...] One of the survey’s questions was: ``Do you support or oppose requiring a background check on all gun buyers?'' 86 percent of respondents said support, 7 percent said oppose, and 8 percent said unsure, according to the Giffords report on the poll. [...] A March 2019 poll conducted by Quinnipiac University found that 87 percent of national gun owners support ``requiring background checks for all gun owners''. The poll surveyed 1,120 national voters and had a margin of error of plus or minus 3.4 percentage points.

\subsubsection*{Evidence 2}
Iowa voters will decide in 2022 whether to add a state Constitution amendment affirming the right to bear arms. Support for the amendment has fallen along party lines. Opposing Democrats refer to a poll that says an overwhelming number of even gun owners support background checks for gun buyers.

\subsubsection*{Uncertainty Explanation}
The parts of the evidence with the most impact on the AI system’s certainty in the prediction are: \\
1. The evidence in Evidence 1, ``86\% of Americans and 82\% of gun owners support requiring all gun buyers to pass a background check'', aligns perfectly with the claim ``86\% of Americans and 18\% of gun owners support requiring all gun buyers to pass a background check''. This consistency increases the certainty that the claim is true. \\
2. The agreement between the phrase ``gun buyers'' in the claim and ``gun'' in Evidence 1 indicates that the focus on background checks applies uniformly across gun-related transactions and increases the certainty in the prediction. \\
3. Lastly, the phrase ``support'' in the claim and ``owners'' in Evidence 1, while seemingly unrelated in terms of direct content, contribute to the overall coherence of the argument. The emphasis on the level of support among gun owners corroborates the claim's specific statistic, and increases certainty about the alignment between the claim and the provided evidence.

\subsubsection*{Verdict Explanation}
The parts of the evidence with the most impact on the AI system’s predicted verdict are: \\
1. Evidence 1 directly supports the claim, stating that ``86\% of Americans and 82\% of gun owners support requiring all gun buyers to pass a background check, no matter where they buy the gun and no matter who they buy it from''. The poll details show that when asked, ``Do you support or oppose requiring a background check on all gun buyers?'', 86\% responded ``support''. This aligns exactly with the percentages in the claim. \\
2. Evidence 1 cites a Quinnipiac University poll reporting that ``87 percent of national gun owners support requiring background checks for all gun owners'', which reinforces the claim’s assertion of broad support among both the general public and gun owners. \\
3. Evidence 2 does not give exact percentages but echoes the same conclusion. It states that ``an overwhelming number of even gun owners support background checks for gun buyers'', which is consistent with the data presented in Evidence 1. While it is less specific, it corroborates the general finding of widespread support.

\subsubsection*{No Explanation}
The AI system has judged this claim to be True, with 76\% certainty.

\subsection{Claim 2 (False, Correct, High Certainty)}
\textbf{Claim:} A satanic-themed hotel is opening in Texas.

\noindent \textbf{AI Prediction:} False

\noindent \textbf{AI Certainty:} 67\%

\subsubsection*{Evidence 1}
Baphomet bedside buddies, upside down crosses, spewing Satan sinks and a devilish welcome, one Facebook user claimed a satanic hotel would soon be opening in Plano, Texas. But don't pull out the rosaries and holy water yet, the hotel is not actually real. On Feb. 23, a Facebook user claimed that a satanic hotel was supposedly opening in an abandoned office building in the heart of the Downtown Plano Art District. [...] But the Baphomet-themed hotel won’t be coming anywhere near Plano. The images were created using an AI art platform. According to The Buzz, an Instagram account posted photos showing how AI art can be made using a specific algorithm. In this case, a satanically-spooky hotel.

\subsubsection*{Evidence 2}
A video shared on Facebook allegedly shows images of a satanic-themed hotel that is set to open in Plano, Texas. [...] A Facebook video allegedly shows photos of a Satanic-themed hotel that is set to open in Plano on June 6. The post shares a photo of a red bedroom with a goat-like figure. ``This is a satanic hotel located in Plano, Texas'' the caption reads. ``This brand new hotel will open June 6 at 6pm (666).'' The claim is fabricated. Check Your Fact found no credible news reports about a satanic-themed hotel opening in Plano, Texas. The photo originates from Ink Poisoning, an apparel brand. The company posted the images on Facebook with a disclaimer stating the images are ``AI concept art created by us''.

\subsubsection*{Uncertainty Explanation}
The parts of the evidence with the most impact on the AI system’s certainty in the prediction are: \\
1. The evidence in Evidence 1, ``Baphomet bedside buddies, upside down crosses, spewing Satan sinks and a devilish welcome, one Facebook user claimed a satanic hotel would soon be opening in Plano, Texas'', is somewhat related to the claim ``A satanic-themed hotel is opening in Texas''. This reduces the AI system’s certainty that the claim is false.\\
2. The evidence in Evidence 1, ``the hotel is not actually real'', is in direct contradiction to the claim ``A satanic-themed hotel is opening in Texas''. This statement increases the certainty that the claim is false by clarifying that the images and descriptions of the hotel are merely AI-generated concepts and not an actual establishment.\\
3. The agreement between the evidence in Evidence 1, that  ``the images were created using an AI art platform'', and the phrase ``supposedly opening'' increase AI system uncertainty that the claim is false.

\subsubsection*{Verdict Explanation}
The parts of the evidence with the most impact on the AI system’s predicted verdict are: \\
1. Evidence 1 makes clear that the idea of a satanic-themed hotel opening in Plano, Texas, comes from a social media post and is not real. It explains that a Facebook user claimed ``a satanic hotel would soon be opening in Plano, Texas'', but the article quickly clarifies that ``the hotel is not actually real''. The supposed images of the ``Baphomet bedside buddies, upside down crosses, spewing Satan sinks and a devilish welcome'' were not photos of a real hotel but instead ``created using an AI art platform''. \\
2. The Buzz further explained that the Instagram account that posted them did so to show ``how AI art can be made using a specific algorithm''. This confirms the images were fictional concept art, not evidence of an actual hotel. \\
3. Evidence 2 supports this conclusion. It describes how a Facebook video claimed ``a satanic-themed hotel… is set to open in Plano on June 6'' and even added details like ``open June 6 at 6pm (666)''. However, the evidence directly states that ``the claim is fabricated'' and that ``no credible news reports'' support the idea of such a hotel opening. It also identifies the true origin of the images: they came from Ink Poisoning, an apparel brand, which shared them with a disclaimer noting they were ``AI concept art created by us''.

\subsubsection*{No Explanation}
The AI system has judged this claim to be False, with 67\% certainty.

\subsection{Claim 3 (True, Correct, Low Certainty)}

\textbf{Claim:} A scone can equal a third of one’s recommended daily calories.

\noindent \textbf{AI Prediction:} True

\noindent \textbf{AI Certainty:} 35\%

\subsubsection*{Evidence 1}
An average scone equals a fifth of recommended daily calories for females, and a sixth for males. The number of calories tends to be proportionate to the size of a scone rather than the luxuriousness of its filling.

\subsubsection*{Evidence 2}
On 11 February 2019, a headline of the Irish News and the Belfast Telegraph stated that ``One large scone can equal a third of recommended daily calories''. Both articles used a Press Association report that referred to ``a survey''. [...] The report states that ``the highest calorie scone, without the addition of spread or jam, provides over a third (38\%) of the recommended daily calorie intake''. [...] For people aged 19 to 64, the Government Dietary Recommendations suggest a daily intake of 2,000kcal for females and 2,500kcal for males. The largest scone of the survey, then, indeed equals a third of recommended calories; taking up 37.8\% of a daily female diet, and 30.2\% of a daily male diet.

\subsubsection*{Uncertainty Explanation}
The parts of the evidence with the most impact on the AI system’s certainty in the prediction are:
1. The evidence in Evidence 1, ``An average scone equals a fifth of recommended daily calories for females, and a sixth for males'', aligns with the claim ``A scone can equal a third of one’s recommended daily calories''. Although the specific fractions differ, the core message that a scone can be a substantial contributor to daily calorie consumption is consistent, increasing the certainty that the claim is true.\\
2. The evidence in Evidence 2, ``without the addition of spread or jam, provides over a third (38\%) of the recommended daily calorie intake'', corroborates the claim by providing a concrete example where a larger scone meets or exceeds the claim's assertion. This specific instance strengthens the certainty that the claim is true.\\
3. The evidence in Evidence 2, ``the largest scone of the survey, then, indeed equals a third of recommended calories; taking up 37.8\% of a daily female diet, and 30.2\% of a daily male diet'', further reinforces the certainty that the claim is true by offering precise figures that closely match the claim's statement.

\subsubsection*{Verdict Explanation}
The parts of the evidence with the most impact on the AI system’s predicted verdict are:
1. Evidence 1 states that an average scone provides ``a fifth of recommended daily calories for females, and a sixth for males''. This shows that even a regular-sized scone contains a significant share of daily energy needs.\\
2. Evidence 1 also emphasizes that calorie content is more closely tied to the size of the scone than to its filling, suggesting that larger scones will contribute an even higher proportion of daily calories.\\
3. Evidence 2 confirms that in some cases a single scone can equal about a third of recommended calories. It explains that ``the highest calorie scone, without the addition of spread or jam, provides over a third (38\%) of the recommended daily calorie intake''. This is further broken down to ``37.8\% of a daily female diet, and 30.2\% of a daily male diet'', based on standard intake levels of 2,000 kcal for women and 2,500 kcal for men.

\subsubsection*{No Explanation}
The AI system has judged this claim to be True, with 35\% certainty.

\subsection{Claim 4 (False, Correct, Low Certainty)}

\textbf{Claim:} A windmill could spin until it falls apart and never generate as much energy as was invested in building it.

\noindent \textbf{AI Prediction:} False

\noindent \textbf{AI Certainty:} 31\%

\subsubsection*{Evidence 1}
No, a professor didn't say a windmill will `never generate' the energy invested in building it [...] Now, we have a meme roaming social media that claims Canadian author and professor Thomas Homer-Dixon said that windmills will ``never'' generate as much energy as was expended to build the structures. [...] The meme does get a part of the quote right, but it strategically leaves out large swaths of information in order to give people the impression that all windmills never expend the amount of energy that was invested in it, which is not accurate. The meme's omitted section explains that, while poorly placed windmills may never generate enough energy payback, a good wind site could generate it in three years or less.

\subsubsection*{Evidence 2}
Factually Inaccurate: Analyzing the energy used in the production of wind turbines does not show that it's greater than the electricity produced over the turbine's working lifetime.

\subsubsection*{Uncertainty Explanation}
The parts of the evidence with the most impact on the AI system’s certainty in the prediction are:\\
1. The evidence in Evidence 1, ``No, a professor didn't say a windmill will `never generate' the energy invested in building it'', directly contradicts the claim's assertion that windmills will never generate sufficient energy. This contradiction increases the AI system’s certainty that the claim is False.\\
2. The evidence in Evidence 1, ``will `never generate' the energy invested'', aligns with the part of the claim ``spin until it falls apart and never generate as much energy as was invested'', reducing the AI system’s certainty that the claim is False. \\
3. The evidence in Evidence 1, ``expended to'', aligns with the phrase ``invested in building'' in the claim, indicating agreement on the energy expenditure aspect, reducing the certainty that the claim is False.

\subsubsection*{Verdict Explanation}
The parts of the evidence with the most impact on the AI system’s predicted verdict are:\\
1. Evidence 1 explains that the claim comes from a misrepresented quote. The meme states that windmills ``will never generate as much energy as was expended to build the structures'', but the original quote clarifies that ``while poorly placed windmills may never generate enough energy payback, a good wind site could generate it in three years or less''. This shows the claim exaggerates and omits important context.\\
2. The same evidence highlights that well-sited windmills can generate enough energy to offset their construction cost quickly, making the claim that all windmills `never generate' enough energy inaccurate. The phrase ``a good wind site could generate it in three years or less'' directly contradicts the claim.\\
3. Evidence 2 reinforces this, stating that ``analyzing the energy used in the production of wind turbines does not show that it’s greater than the electricity produced over the turbine's working lifetime''. This confirms that, over time, turbines produce more energy than was invested in building them.

\subsubsection*{No Explanation}
The AI system has judged this claim to be False, with 31\% certainty.

\subsection{Claim 5 (True, Incorrect, High Certainty)}

\textbf{Claim:} All tea bags contain harmful microplastics.

\noindent \textbf{AI Prediction:} True

\noindent \textbf{AI Certainty:} 73\%

\subsubsection*{Evidence 1}
At ORGANIC INDIA, we believe in plastic-free tea bags to protect your body and the earth; and preserve the pure, organic tea experience that you deserve. Unfortunately, the industry norm is to include plastic in the mesh tea bag, which releases billions of microplastics and nanoplastics into your infusion. Consuming microplastics, as you can imagine, can have negative impacts on your health and on the earth. [...] Yes, but not all tea bags. The vast majority of brands on the shelves have mesh tea bags that are composed of 20-30\% plastic. According to studies, a single standard tea bag releases 11.6 billion microplastics and 3.1 billion nanoplastics into every cup of tea. Consumers are increasingly becoming aware of this fact and opting either for loose teas, or selecting tea brands that have no plastic in the teabag. ORGANIC INDIA tea bags do not contain plastic, and are made instead of unbleached filter paper that is verified home compostable, biodegradable, Non-GMO, allergen-free and chemical-free. [...] In short, yes. Microplastics in tea bags can be harmful not only for the environment, but for human health as well. Microplastics may enter the human system through ingestion, inhalation or dermal absorption. Studies have shown that microplastics have potential to cause damage to the cells in the quantities currently found in food and beverages like tea.

\subsubsection*{Evidence 2}
Therefore, the use of plastic in tea bags has come under scrutiny due to the potential release of microplastics when exposed to heat. As we mentioned earlier, many tea bag manufacturers use nylon or polyethylene terephthalate (PET) in their tea bags, which are not biodegradable and can release harmful microplastics when exposed to heat. These microplastics can end up in your drink, and studies have shown that they may pose a risk to human health. [...] Another study published in the journal Environmental Science \& Technology in 2020 found that plastic tea bags could release up to 16 micrograms of microplastics per liter of tea when brewed at a temperature of 95°C. These studies highlight the importance of choosing plastic-free tea bags to minimize exposure to potentially harmful plastics. [...] Yes, some tea bags can contain microplastics, particularly those made with plastic materials like nylon or polypropylene. However, The Tea Spot's sachets are made from sugar cane fibers, which are biodegradable, compostable, and completely free from petroleum-based plastic. Therefore, you can rest assured that when you brew our tea, you're not consuming any harmful plastics or microplastics.

\subsubsection*{Uncertainty Explanation}
The parts of the evidence with the most impact on the AI system’s certainty in the prediction are:\\
1. The statement in Evidence 2, ``Therefore, the use of plastic in tea bags has come under scrutiny due to the potential release of microplastics when exposed to heat'', does not support the blanket statement that ``All tea Bags contain harmful microplastics''. This disagreement reduces the AI system certainty that the claim is True.\\
2. The statement in Evidence 1, ``Yes, but not all tea bags. The vast majority of brands on the shelves have mesh tea bags that are composed of 10-30\% plastic'', contradicts the claim by indicating that not all tea bags contain harmful microplastics, reducing certainty that the claim is True.\\
3. The statement in Evidence 1, ``ORGANIC INDIA tea bags do not contain plastic, and are made instead of unbleached filter paper that is verified home compostable, biodegradable, Non-GMO, allergen-free and chemical-free'', undermines the claim by providing a specific example of a brand that produces tea bags without harmful microplastics, further reducing certainty that the claim is True.

\subsubsection*{Verdict Explanation}
The parts of the evidence with the most impact on the AI system’s predicted verdict are:\\
1. Evidence 1 states that ``the vast majority of brands on the shelves have mesh tea bags that are composed of 20–30\% plastic''. This shows that most commercially available tea bags contain plastic, making the exposure to microplastics nearly universal for consumers using standard tea bags.\\
2. Evidence 1 notes that ``a single standard tea bag releases 11.6 billion microplastics and 3.1 billion nanoplastics into every cup of tea'', and Evidence 2 confirms that materials like nylon or PET ``can release harmful microplastics when exposed to heat''. Both sources directly link the presence of plastic in tea bags to microplastic contamination in brewed tea.\\
3. Evidence 1 explains that microplastics ``can be harmful not only for the environment, but for human health as well'' and ``may cause damage to the cells in the quantities currently found in food and beverages like tea''. Evidence 2 echoes this, stating that microplastics ``may pose a risk to human health''.

\subsubsection*{No Explanation}
The AI system has judged this claim to be True, with 73\% certainty.

\subsection{Claim 6 (False, Incorrect, High Certainty)}

\textbf{Claim:} Lagan Valley is one of the least wooded areas in Northern Ireland, particularly with regards to ancient woodland.

\noindent \textbf{AI Prediction:} False

\noindent \textbf{AI Certainty:} 78\%

\subsubsection*{Evidence 1}
The Lagan Valley (Irish: Cluain an Lagáin, Ulster Scots: Glen Lagan) is an area of Northern Ireland between Belfast and Lisburn. The Lagan is a famous river that flows into Belfast Lough. For a section, the river forms part of the border between the counties of Antrim and Down. It has a number of interesting features including a towpath which runs alongside the River Lagan. The towpath is popular with walkers, runners, cyclists, dog owners etc. It is a very scenic and peaceful area and is ideal for walking, cycling etc. The towpath begins in the Stranmillis area of south Belfast and runs all the way to Lisburn. The cycle route forms part of National Cycle Route 9. There are a number of ``off route'' mountain bike trails along the towpath. Lagan Valley is an Area of Outstanding Natural Beauty (AONB). The AONB was established in 1965 and the greater part of it lies within the Greater Belfast area. [...] Belvoir Park Forest Education Centre 

\subsubsection*{Evidence 2}
Therefore, it is fair to say that Lagan Valley is one of the least wooded areas in NI. [...] Lisburn \& Castlereagh is the least densely wooded council (3.7\%) in Northern Ireland, followed by Armagh, Banbridge \& Craigavon (4.0\%) and Antrim \& Newtownabbey (4.4\%). [...] Based on all this, Ms Eastwood’s claim that ``Lagan Valley is one of the least wooded areas in NI'' is backed by evidence – although the precise figures are available broken down by council areas, rather than Westminster/Assembly constituencies.

\subsubsection*{Uncertainty Explanation}
The parts of the evidence with the most impact on the AI system’s certainty in the prediction are:\\
1. The evidence in Evidence 1, ``Lagan Valley is an Area of Outstanding Natural Beauty (AONB)'', does not directly address the claim about the level of woodlands in the area, reducing the certainty of the AI system’s prediction.\\
2. The evidence in Evidence 2, ``Therefore, it is fair to say that Lagan Valley is one of the least wooded areas in NI'', directly supports the claim that Lagan Valley is one of the least wooded areas in Northern Ireland, particularly with regards to ancient woodland. This agreement reduces the certainty that the claim is False.\\
3. The evidence in Evidence 2 that ``Lisburn \& Castlereagh is the least densely wooded council (3.7\%) in Northern Ireland'' aligns with the evidence in Evidence 1 that the Lagan Valley ``is an area of Northern Ireland between Belfast and Lisburn''. This reduces the certainty that the claim is False.

\subsubsection*{Verdict Explanation}
The parts of the evidence with the most impact on the AI system’s predicted verdict are:\\
1. Evidence 1 describes Lagan Valley as ``an Area of Outstanding Natural Beauty (AONB)'' and mentions that it ``has a number of interesting features including a towpath which runs alongside the River Lagan''. It also emphasizes that the area is ``very scenic and peaceful'' and ``ideal for walking, cycling etc.'', suggesting the presence of natural green spaces rather than a sparsely wooded landscape.\\
2. Evidence 2 notes that ``Lisburn \& Castlereagh is the least densely wooded council (3.7\%) in Northern Ireland, followed by Armagh, Banbridge \& Craigavon (4.0\%) and Antrim \& Newtownabbey (4.4\%)''. It adds that Ms Eastwood’s claim that ``Lagan Valley is one of the least wooded areas in NI'' is ``backed by evidence – although the precise figures are available broken down by council areas, rather than Westminster/Assembly constituencies'', showing that the claim is not directly supported for Lagan Valley itself.
3. Neither piece of evidence provides specific data about ancient woodland in Lagan Valley. The lack of precise figures for the area means the claim about ancient woodland cannot be verified.

\subsubsection*{NoExplanation}
The AI system has judged this claim to be False, with 78\% certainty.

\subsection{Claim 7 (True, Incorrect, Low Certainty)}

\textbf{Claim:} Cristiano Ronaldo was the only European soccer captain who did not wear a rainbow armband at Euro 2020.

\noindent \textbf{AI Prediction:} True

\noindent \textbf{AI Certainty:} 29\%

\subsubsection*{Evidence 1}
Cristiano Ronaldo is refusing to wear the ``One Love'' armband, which advocates for the LGBTQ+ community, during the 2022 World Cup. The world-renowned soccer player, Cristiano Ronaldo, revealed in a press conference that he will not be participating in wearing the ``One Love'' armband. The armband consists of a heart with a rainbow inside of it and the number one. This ``One Love'' band is being used to advocate inclusivity, being that the 2022 World Cup is being held in Qatar, a country where same-sex relationships are illegal. [...] Players who wear the armband are threatened with receiving a yellow card, according to FIFA’s regulations. This yellow card can result in a suspension from the games if the player were to receive another yellow card during the game. Obviously, any mention of social issues is discouraged on the field. Due to the consequences, none of the captains from any of the other national teams have worn the armband, but none have faced backlash like Ronaldo has.

\subsubsection*{Evidence 2}
There is evidence that three team captains — from Germany, England, and Denmark — wore rainbow or pro-LGBTQ+ armbands during this tournament, which included 24 teams. Germany’s captain, Manuel Neuer, was briefly investigated by the Union of European Football Associations for his rainbow armband. Cristiano Ronaldo was among many other captains who wore normal armbands.

\subsubsection*{Uncertainty Explanation}
The parts of the evidence with the most impact on the AI system’s certainty in the prediction are:\\
1. The evidence in Evidence 1, ``team captains — from Germany, England, and Denmark — wore rainbow or pro-LGBTQ+ armbands during this tournament'', is not consistent with the part of the claim ``Cristiano Ronaldo was the only European soccer captain who did not wear a rainbow armband at Euro 2020''. This indicates that there were multiple captains who did not wear the armband, not just Ronaldo and reduces AI system certainty that the claim is True.\\
2. The evidence in Evidence 1, ``Cristiano Ronaldo was among many other captains who wore normal armbands'', is not consistent with the part of the claim ``Cristiano Ronaldo was the only European soccer captain.'', implying he was not the only captain without a rainbow armband, reducing certainty that the claim is True.\\
3. The evidence in Evidence 1, ``during this tournament'', is not consistent with the part of the claim ``Euro 2020'', as it refers to a different event. This discrepancy highlights that the evidence discusses a different tournament, possibly the 2022 World Cup, rather than Euro 2020 and reduces certainty that the claim is True.

\subsubsection*{Verdict Explanation}
The parts of the evidence with the most impact on the AI system’s predicted verdict are:\\
1. Evidence 1 notes that Ronaldo ``is refusing to wear the `One Love' armband, which advocates for the LGBTQ+ community,'' and describes how he revealed in a press conference that he ``will not be participating in wearing the `One Love' armband.'' The evidence emphasizes that while ``players who wear the armband are threatened with receiving a yellow card'', Ronaldo’s decision not to wear it drew notable attention, suggesting he stood out among team captains. The passage also clarifies that ``none of the captains from any of the other national teams have worn the armband'' at other tournaments, highlighting Ronaldo’s high-profile refusal.\\
2. Evidence 2 provides additional support, stating that ``three team captains — from Germany, England, and Denmark — wore rainbow or pro-LGBTQ+ armbands during this tournament''. It also specifically mentions that ``Cristiano Ronaldo was among many other captains who wore normal armbands'', directly confirming that he did not wear the rainbow armband while other European captains did.

\subsubsection*{No Explanation}
The AI system has judged this claim to be True, with 29\% certainty.

\subsection{Claim 8 (False, Incorrect, Low Certainty)}

\textbf{Claim:} The top issue for college students is the economy.

\noindent \textbf{AI Prediction:} True

\noindent \textbf{AI Certainty:} 49\%

\subsubsection*{Evidence 1}
``The everyday concerns of Americans are the economy — even young people that we work with at Young America's Foundation, our polling shows nationwide, their top issue for college students is the economy'', Walker said. [...] Is Walker right that ``The top issue for college students is the economy''? [...] The economy was the top issue for college students who participated in the group's poll. Wisconsin pollster Charles Franklin said the polling company is reputable and the questions were overall evenhanded.

\subsubsection*{Evidence 2}
The midterm election is right around the corner and the economy and abortion rights are the top issues of concern for college students according to a new survey by BestColleges.com. These issues are also ranked as the most important for Americans overall. BestColleges.com surveyed over 1000 students about their political beliefs and plans to vote. Thirty-seven percent (37\%) of those surveyed noted the economy, inflation, and employment as the most important issues in this election. Students are worried about finding jobs, their future finances, and the growing cost of everything from housing to peanut butter. White college men, in particular, are overwhelmingly concerned about the economy, employment, and inflation, with almost half (49\%) selecting these three issues as their top political issues — a significantly higher percentage than other demographic groups surveyed. Abortion rights, which have been the subject of heated national and local political campaigns, were ranked as a top concern by college students with 33\% of those surveyed ranking these rights as a top issue. College women were particularly concerned about legislation having an impact on their bodies. While college men chose the economy as their top issue (44\%), abortion rights were the most important issue for college women (43\%), with women of color expressing the deepest concern. Only 21\% of college men considered abortion rights an important issue. For college men, the economy, gun policy, healthcare, racial and ethnic inequality, and climate change all came out ahead of abortion rights as leading issues.

\subsubsection*{Uncertainty Explanation}
The parts of the evidence with the most impact on the AI system’s certainty in the prediction are:\\
1. The evidence in Evidence 1, ``The economy was the top issue for college students who participated in the group’s poll'', is consistent with the claim ``The top issue for college students is the economy'', reducing AI system certainty that the claim is False.
2. The evidence in Evidence 2, ``abortion rights were the most important issue for college women (43\%)'' contrasts with the claim, increasing certainty that the claim is False.\\
3. The statement in Evidence 1, ``The everyday concerns of Americans are the economy'', is consistent with the part of Evidence 2, ``Students are worried about finding jobs, their future finances, and the growing cost of everything'',  reducing AI system certainty that the claim is False.

\subsubsection*{Verdict Explanation}
The parts of the evidence with the most impact on the AI system’s predicted verdict are:\\
1. Evidence 1 reports that, according to polling by Young America’s Foundation, ``the economy was the top issue for college students who participated in the group’s poll''. While this appears to support the claim, the evidence also raises the question: ``Is Walker right that `The top issue for college students is the economy'?'' suggesting uncertainty about whether this finding is representative of all college students.\\
2. Evidence 2 provides more detailed data, showing that college students are concerned about multiple issues. It states that ``the economy and abortion rights are the top issues of concern for college students'', with 37\% ranking the economy and 33\% ranking abortion rights as their primary concern. The evidence further breaks this down by gender: ``college men chose the economy as their top issue (44\%)'', whereas ``abortion rights were the most important issue for college women (43\%), with women of color expressing the deepest concern''. This demonstrates that the economy is not universally the top issue across all college students.

\subsubsection*{No Explanation}
The AI system has judged this claim to be False, with 49\% certainty.



\end{document}